\documentclass[aps,prb,amsmath,amssymb,footinbib,showpacs,twocolumn]{revtex4-1}
\usepackage{amsmath}
\usepackage{amssymb}
\usepackage{amsthm}
\usepackage{setspace}
\usepackage{graphicx}
\usepackage{braket}
\usepackage{mathrsfs}
\usepackage{physics}
\usepackage{float}
\usepackage[colorlinks = true,linkcolor = red,urlcolor  = blue,citecolor = blue,anchorcolor = blue]{hyperref}
\usepackage{bm}
\usepackage{xcolor}

\begin{document}
	\title{Three-terminal nonlocal conductance in Majorana nanowires: Distinguishing topological and trivial in realistic systems with disorder and inhomogeneous potential}
	\author{Haining Pan}
	\author{Jay D. Sau}
	\author{S. Das Sarma}
	\affiliation{Condensed Matter Theory Center and Joint Quantum Institute, Department of Physics, University of Maryland, College Park, Maryland 20742, USA}
	
		\begin{abstract}
	We develop a theory for the three-terminal nonlocal conductance in Majorana nanowires as existing in the superconductor-semiconductor hybrid structures in the presence of superconducting proximity, spin-orbit coupling, and Zeeman splitting.  The key question addressed is whether such nonlocal conductance can decisively distinguish between trivial and topological Majorana scenarios in the presence of chemical potential inhomogeneity and random impurity disorder. We calculate the local electrical as well as nonlocal electrical and thermal conductance of the pristine nanowire (good zero-bias conductance peaks), the nanowire in the presence of quantum dots and inhomogeneous potential (bad zero-bias conductance peaks), and the nanowire in the presence of large disorder (ugly zero-bias conductance peaks). The local conductance by itself is incapable of distinguishing the trivial states from the topological states since zero-bias conductance peaks are generic in the presence of disorder and inhomogeneous potential. The nonlocal conductance, which in principle is capable of providing the bulk gap closing and reopening information at the topological quantum phase transition, is found to be far too weak in magnitude to be particularly useful in the presence of disorder and inhomogeneous potential. Therefore, we focus on the question of whether the combination of the local, nonlocal electrical, and thermal conductance can separate the good, bad, and ugly zero-bias conductance peaks in finite-length wires. Our paper aims to provide a guide to future experiments, and we conclude that a combination of all three measurements would be necessary for a decisive demonstration of topological Majorana zero modes in nanowires--- positive signals corresponding to just one kind of measurements  are likely to be false positives arising from disorder and inhomogeneous potential.
	\end{abstract}
	\date{\rm\today}
	\maketitle
	
	\section{Introduction}\label{sec:introduction}
	The experimental search for Majorana zero modes~\cite{kitaev2003faulttolerant,dassarma2005topologically,nayak2008nonabelian} (MZM) in the superconductor-semiconductor (SC-SM) hybrid devices~\cite{sau2010generic,sau2010nonabelian,oreg2010helical,lutchyn2010majorana} has found many theoretically-predicted signatures of the MZM, especially the zero bias conductance peak (ZBCP) in the normal-to-superconductor (NS) tunneling spectroscopy.~\cite{mourik2012signatures,das2012zerobias,deng2012anomalous,churchill2013superconductornanowire,finck2013anomalous, deng2016majorana,albrecht2016exponential,kammhuber2016conductance,kammhuber2017conductance,nichele2017scaling,zhang2017ballistic,chen2017experimental,vaitiekenas2018effective,moor2018electric,gul2018ballistic,zhang2018quantized,bommer2019spinorbit,grivnin2019concomitant,chen2019ubiquitous,anselmetti2019endtoend,menard2020conductancematrix,yu2021nonmajorana,puglia2020closing} In fact, even the putative Majorana conductance quantization has been observed in some of the two-terminal NS tunneling measurements.~\cite{zhang2018quantized} However, these experiments with one normal lead and one superconductor lead (i.e., two-terminal device) have not yet confirmed other hallmarks of MZM, including the robust stability of ZBCP in an extended region over the various gate voltages and magnetic field,~\cite{chen2017experimental,chen2019ubiquitous,yu2021nonmajorana} the closing and reopening of the bulk superconductor gaps,~\cite{huang2018metamorphosis} and the growing Majorana oscillation with the increasing magnetic field.~\cite{dassarma2012splitting} In fact, it is correct to say that neither any direct signature for a topological quantum phase transition nor any property of the bulk (i.e., away from the tunnel contacts) has yet been observed experimentally, leading to justified skepticism about the existence of topological anyonic physics in the system in spite of the reported abundance of ZBCPs in many experiments.  These ZBCPs may just be ``bad" or ``ugly" nontopological zero bias peaks arising from disorder in the system and not the hoped-for ``good" ZBCPs arising from topological MZMs in pristine systems.~\cite{pan2020generic}
	
	For example, the Majorana oscillation--- though never directly observed in experiments--- is an established hallmark of the real topological MZM. There are several reasons for the absence of the Majorana oscillations in the local conductance,\cite{liu2012zerobias,bagrets2012class,pikulin2012zerovoltage,lee2012zerobias,cayao2015sns,clarke2017experimentally,liu2017andreev,reeg2018zeroenergy,fleckenstein2018decaying,pan2020physical} e.g., the superconductor bulk gap collapses too early below the putative topological quantum phase transition (TQPT), the self-energy effect is too strong so that it suppresses the amplitude of the oscillation, the omnipresent trivial ZBCP in class D ensemble due to the disorder in the nanowire,~\cite{mi2014xshaped,pan2020generic} etc. {The total absence of MZM oscillations in all experiments reporting ZBCPs, however, remains a serious problem, particularly since the nanowires used in the experiments are generally rather short, and some oscillatory behavior is expected in the conductance even in the presence of self-energy effects.}

	Therefore, the measurement of local conductance through NS tunneling, just by itself, is useless in determining the good, bad, and ugly ZBCP, because many accidental Andreev bound states (ABS) have nearly zero energy in the nanowire.~\cite{huang2018metamorphosis} Even if the disorder is weak, the problem of ABS itself is already notoriously serious.~\cite{liu2017andreev} Without any bulk information (e.g., the bulk gap as a function of {Zeeman} field), the local conductance cannot distinguish whether the Andreev state is perfect or not (the Majorana bound state, MBS, is simply a ``perfect" ABS, being a precise midgap zero energy state, the MZM), and thus whether the system is in the topological regime or not. What is even worse, the nanowire system is often highly disordered, and the strong disorder creates ubiquitous trivial ``class D" ZBCP that can even appear to be quantized through careful fine-tuning of system parameters (e.g., various gate voltages).~\cite{pan2020generic} Therefore, the local conductance measurement may be necessary, but is by no means sufficient in distinguishing ABS from MBS. Even after the manifestation of a quantized ZBCP, additional experimental evidence is necessary to establish the nonlocal nature of the system through a suitable bulk measurement beyond two-terminal NS tunneling spectroscopy.
	
	One crucial drawback of the two-terminal ZBCP tunneling {measurements} is that the experiment does not probe bulk properties and cannot therefore provide any direct information on nonlocality, which is the characteristic signature of topological MZMs. There have been suggestions and efforts to circumvent this problem by doing tunneling measurements from both ends and looking for correlations in the ZBCPs produced from the two ends and/or carrying out local conductance measurements using tunnel probes throughout the nanowire.  But, as already known, such inherently two-terminal measurements are incapable of providing information about the bulk gap closure and reopening, which, along with the appearance of a quantized ZBCP, is the hallmark of MZM physics.  In the current work, we analyze a promising candidate, a direct measurement of the three-terminal nonlocal conductance, as a possible probe for MZMs--- this measurement is in principle capable of providing the bulk gap information missing in the two-terminal NS tunneling measurement.
	
	A conceptually clean way to detect the bulk property is the nonlocal conductance measuring the tunneling conductance from one end to the other. Several experiments have already performed measurements in three-terminal devices,~\cite{menard2020conductancematrix,puglia2020closing} though they are incomplete and inconclusive, because only a few situations are studied, and no complete picture has emerged with characteristic MZM signatures. Given the potential importance of nonlocal three-terminal experiments in the MZM observation, we develop a complete theory for nonlocal conductance in nanowires including effects of potential inhomogeneity (e.g., quantum dots or a smoothly varying chemical potential) and random disorder, comparing these realistic results with the pristine situation. Our work, comparing ``good'' (i.e., pristine), ``bad" (i.e., inhomogeneous potential and quantum dots), and ``ugly" (i.e. random disorder) cases,~\cite{pan2020physical} should serve as a guide for forthcoming nonlocal conductance measurements. We already know that disorder can create accidental almost-quantized ZBCP on its own in the local conductance,~\cite{pan2020generic,pan2020physical} but the behavior of nonlocal conductance in the presence of inhomogeneous potential and random disorder has not been studied systematically in the literature, and therefore, it is unclear at this stage whether the bulk signatures in the nonlocal conductance of pristine systems survive inhomogeneous potential and/or disorder.  Our goal is to carefully study effects of inhomogeneous background potential and random disorder on the nonlocal conductance of Majorana nanowires in order to help the interpretation of future experiments.
	
	In principle, the nonlocal conductance should be decisive in establishing the existence or not of MZMs since it contains the bulk information and thus can identify the proximity gap as well as its closing and reopening associated with the TQPT: The conductance is vanishingly small above the parent SC gap and below the proximity gap. Therefore, we may directly see where the bulk gap is and how the bulk gap closes (and then reopens) as the magnetic field increases. However, the measurement of nonlocal conductance poses a problem that the signal is too weak since the measurement is nonlocal (and thus of higher order in strength). In this work, we will show explicitly how small the nonlocal conductance could be compared to the local conductance, especially in the presence of disorder, which usually dominates in experiments. Therefore, one should be careful not to confuse the background noise as the small signal of nonlocal conductance. Our work provides a quantitative guide on the expected signal strength of nonlocal conductance (and indeed it is rather small, and noise may very well turn out to be a serious problem in three-terminal measurements).
	
	However, in our theoretical work, we have the advantage that we can directly calculate the topological invariant (TI)~\cite{fulga2011scattering} to identify the topological property. This is obviously something which is impossible in experiments--- in our work we explicitly know by construction whether a particular calculated conductance feature, no matter how weak, arises from topological or trivial physics. We calculate the topological visibility which is only well defined for the gapped state. To determine whether it is gapped or not, we resort to the thermal conductance, which measures the total transmission of electrons and holes. The thermal conductance should peak at a quantized value~\cite{akhmerov2011quantized} at the TQPT. In our work, we can unambiguously separate the three situations, the good, the bad, and the ugly, because we know by construction which results belong to which category. In experiments, however, one only has the final conductance results, where distinguishing between topological (``good") and trivial (``bad" or ``ugly") may be a huge challenge, particularly if the nonlocal conductance signals are so low that they are overwhelmed by noise leading to all three situations producing qualitatively similar conductance.  We already know from the situation in the local conductance, as obtained by two-terminal NS tunneling, even ZBCPs which appear to be quantized may be generically ``bad" or ``ugly", so one should be very careful in interpreting the nonlocal conductance experimental results when they come out. Therefore, as a sequel to Ref.~\onlinecite{pan2020physical} where only the local conductance was considered, we aim to answer in the current paper whether the combination of local conductance, nonlocal conductance, and thermal conductance can decisively distinguish the topological ``good" case from the trivial ``bad" and ``ugly" cases. 
	
	We first present the electrical conductance and thermal conductance of a pristine nanowire (good ZBCP) to provide a general picture of what the nonlocal conductance in the three-terminal device should ideally look like. Next, we consider a more complicated case--- the bad ZBCP in the presence of the quantum dot and an inhomogeneous chemical potential. Then we add disorder to the wire to study the disorder effect, where we find weak disorder preserves the good ZBCP while strong disorder can induce the trivial ugly ZBCP. When disorder is very strong, the whole concept of the topological superconductor is pointless since the SC gap is destroyed leading to Anderson localization. Therefore, the strongly disordered nanowire just breaks into a series of quantum dots, which should be described by the random matrix theory.~\cite{beenakker1997randommatrix,guhr1998randommatrix,brouwer1999distribution,beenakker2015randommatrix,mi2014xshaped,pan2020generic} Finally, we consider a more realistic scenario by combining the quantum dot and the disorder. To correspond to the realistic experimental scenarios, we also consider short wire lengths as well as the magnetic field induced bulk gap collapse situations to investigate realistic complications expected to arise in the experiments.  We present a few sets of the representative results of the good and ugly ZBCP in the main text, and relegate the rest to the Appendix, where most of our detailed results are provided.
	
	The paper is organized as follows. In Sec.~\ref{sec:theory}, we start with the minimal effective model of a pristine nanowire (good ZBCP) and modify each term in the Hamiltonian to study the inhomogeneous potential (bad ZBCP) and the disorder (ugly ZBCP).  We also define the electrical conductance, thermal conductance, and the topological visibility here. In Sec.~\ref{sec:results}, we show several representative results of the good, bad, and ugly ZBCP in both long and short wires with and without the parent SC bulk gap collapse, and critically discuss the efficacy of the combination of the local, nonlocal, and thermal conductance in distinguishing the trivial ABS from topological MBS in Sec.~\ref{sec:discussion}. Our conclusion is in Sec.~\ref{sec:conclusion}. In Appendix~\ref{app:A}, we present extensive results for good, bad, and ugly situations for comparison with future experiments.

	\section{Theory}\label{sec:theory}
	\subsection{Minimal effective model}
	The most general Hamiltonian to describe the one-dimensional SC-SM nanowire model is~\cite{stanescu2013majorana} 
	\begin{equation}\label{eq:H}
		H=H_{\text{SM}}+H_{\text{Z}}+H_{\text{V}}+H_{\text{SC}}+H_{\text{SC-SM}},
	\end{equation}
	where $ H_{\text{SM}} $ describes the SM, $ H_{\text{Z}} $ results from the magnetic field (entering as the Zeeman splitting energy), $ H_{\text{V}} $ accounts for various effects of the gate voltage potentials and disorder, $ H_{\text{SC}} $ describes the SC, and $ H_{\text{SC-SM}} $ is the coupling between SC and SM. This model was introduced in Refs.~\onlinecite{sau2010generic} and~\onlinecite{sau2010nonabelian} but was not systematically studied with the inclusion of various possible $ H_{\text{V}} $ until recently in Ref.~\onlinecite{pan2020physical}. In this paper, we generalize the previous two-terminal local conductance in Ref.~\onlinecite{pan2020physical} to the three-terminal nonlocal conductance as shown in Fig.~\ref{fig:schematic} and critically study the efficacy of three-terminal measurements for a definitive confirmation of topological MBS. Moreover, we theoretically calculate the topological invariant and thermal conductivity to discuss this possibility.
	
	We start with the pristine wire case omitting $ H_{\text{V}} $, which we call the ``good'' ZBCP as in Ref.~\onlinecite{pan2020physical}. The pristine nanowire is then described in the minimal model by a Bogoliubov-de Gennes Hamiltonian $ H_{\text{BdG}}=\frac{1}{2}\int dx \hat{\Psi}(x)^\dagger H_{\text{eff}} \hat{\Psi}(x)$, {where $ \hat{\Psi}(x) =\left(\hat{\psi}_{\uparrow}(x),\hat{\psi}_{\downarrow}(x),\hat{\psi}_{\downarrow}^\dagger(x),-\hat{\psi}_{\uparrow}^\dagger(x)\right)^T$ is a position-dependent Nambu spinor. The retarded Green's function $ G_{R}(\omega) $ for the Bogoliubov quasiparticles in the Nambu basis, which allows us to compute all time-dependent response functions for an effectively noninteracting theory, can be written in terms of the effective Hamiltonian,
	\begin{equation}\label{eq:bdg-pristine}
		H_{\text{eff}}=\left( -\frac{\hbar^2}{2m^*} \partial^2_x -i \alpha \partial_x \sigma_y - \mu \right)\tau_{z} + V_{\text{Z}}\sigma_x + \Sigma(\omega),
	\end{equation}
	where $ G_{R}(\omega)=\qty[H_{\text{eff}}\qty(\omega+i\eta)-(\omega+i\eta)]^{-1} $ ($\eta$ being the standard positive infinitesimal required to ensure causality).}
	Here $ \vec{\bm{\sigma}} $ and $ \vec{\bm{\tau}} $ act on spin and particle-hole space, respectively. The first term in the parenthesis models the SM, which is denoted by $ H_{\text{SM}} $. The second term enters as a Zeeman field term $ H_{\text{Z}}=\frac{1}{2} g \mu_{B} B $, where $ \mu_B $ is the Bohr magneton and $ B $ is the magnetic field applied along the nanowire growth direction ($ x $ direction). The last term $ \Sigma(\omega) $ {arises from SC proximity effect}, which represents the tunneling effect between the SC and the SM as $ H_{\text{SC-SM}} $. By integrating out the SC {degrees} of freedom $ H_{\text{SC}} $,~\cite{stanescu2010proximity} we obtain the self-energy term 
	\begin{equation}\label{eq:SE}
		\Sigma(\omega)=-\gamma\frac{\omega+\Delta_0\tau_x}{\sqrt{\Delta_0^2-\omega^2}},
	\end{equation}
	where $ \gamma $ is the effective SC-SM coupling (tunneling) strength, $ \omega $ is the energy, and $ \Delta_0 $ denotes the parent SC gap. For parameters in Eqs.~\eqref{eq:bdg-pristine} and~\eqref{eq:SE}, unless otherwise specified, we follow the parameters in InSb-Al hybrid nanowire~\cite{lutchyn2018majorana,zhang2018quantized,gul2018ballistic,kammhuber2017conductance,chen2019ubiquitous,woods2019zeroenergy}:  the effective mass $ m^*=0.015 m_e $ ($ m_e $ is the electron rest mass), {Al superconducting gap} $ \Delta_0=0.2 $ meV, chemical potential $ \mu=1 $ meV, Rashba-type spin-orbit coupling (SOC) $ \alpha=0.5 $ eV\AA{} perpendicular to the nanowire growth direction~\cite{woods2019zeroenergy}, SC-SM coupling strength $ \gamma=1 $ meV, temperature $ \tau=0 $, and the length of nanowire $ L=3~\mu$m for what we call the long wire case and $ L=0.5~\mu $m for what we call the short wire case. We mention that most current experiments use short wires, but the exploration of topological MZMs {requires} long wires.

		\subsection{Electrical conductance, thermal conductance and topological invariant}
	\begin{figure}[t]
		\centering
		\includegraphics[width=3.4in]{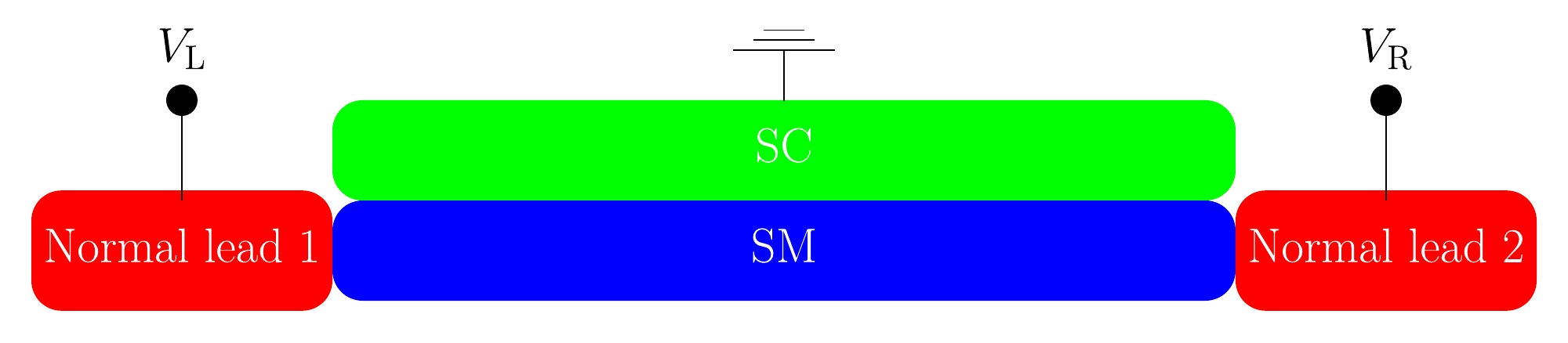}
		\caption{The schematic of the three-terminal device with two normal leads in red and one grounded superconducting lead (SC) in green. The corresponding two-terminal device for measuring the local tunnel conductance simply removes one of the contacts at the wire ends (i.e., either lead 1 or lead 2 is present for two-terminal whereas both are present for three-terminal).}
		\label{fig:schematic}
	\end{figure}
	
	We show the three-terminal device to measure the nonlocal conductance in Fig.~\ref{fig:schematic}. The SM (blue) as a scattering region is covered by a grounded SC lead (green) and separated by two normal leads (red). We apply a bias $ V_L $ on the left normal lead and $ V_R $ on the right normal lead. The normal lead is described by a Hamiltonian in Eq.~\eqref{eq:bdg-pristine} without the SC-related term $ \Sigma(\omega) $. 
	
	Here, we first want to clarify the ambiguity of the two-terminal device and three-terminal device due to the inconsistent nomenclature used in this community: The two-terminal device in Refs.~\onlinecite{moore2018twoterminal,akhmerov2011quantized} is actually the three-terminal device in Refs.~\onlinecite{rosdahl2018andreev,menard2020conductancematrix}. In this paper, we reserve the term ``two-terminal"---one grounded SC lead covering SM plus one normal lead at the end--- to denote the usual local conductance measurement, as initially proposed in theory and performed in experiments. However, the ``three-terminal device"---one grounded SC lead covering SM plus two normal leads at both ends--- measures the nonlocal conductance, i.e., the conductance matrix in Eq.~\eqref{eq:condmat}, which is recently suggested in theory~\cite{moore2018twoterminal,rosdahl2018andreev} and performed in experiments~\cite{menard2020conductancematrix,anselmetti2019endtoend}. In the two-terminal device, we can still perform the correlation measurement~\cite{pan2020physical} by isolating the one lead when measuring from the other. On the contrary, both normal leads are always attached to the SM when measuring conductance in the three-terminal device. The local conductance in two schemes will be slightly different, but the nonlocal conductance is only achievable in the three-terminal device. 
	
	To calculate the electrical and thermal conductance, we use the S matrix formalism with~\cite{datta1995electronic}
	\begin{equation}\label{eq:S}
		S=\mqty*(S_{\text{LL}} & S_{\text{LR}}\\S_{\text{RL}} & S_{\text{RR}}), 
	\end{equation}
	where each block of the scattering matrix connecting lead $ j $ to lead $ i $ is
	\begin{equation}\label{eq:Sij}
		S_{ij}=\mqty*(S_{ij}^{ee} & S_{ij}^{eh} \\ S_{ij}^{he} & S_{ij}^{hh}).
	\end{equation}

	Here the superscript represents a process of transmission from type $ \beta $ to type $ \alpha $, where $ \alpha,\beta=e,h $ stand for the electron and hole. The numerical calculation of the S matrix ~\eqref{eq:S} is performed~\cite{blonder1982transition,buttiker1992scattering,anantram1996current,beenakker2013search,groth2014kwant,sarma2015majorana} with the help of a quantum transport package KWANT.~\cite{groth2014kwant} We discretize the continuum Hamiltonian into {a} finite difference tight-binding model~\cite{dassarma2016how} by setting the lattice constant $ \delta x=10 $ nm. We also model the chemical potential of the lead equaling $ 25 $ meV and the height of barrier at the NS junction equaling $ 5 $ meV~\cite{setiawan2017electron} to match the normal conductance in experiments.~\cite{zhang2018quantized}

	The local conductance ($ G_{\text{LL}} $ and $ G_{\text{RR}} $) and nonlocal conductance ($ G_{\text{LR}} $ and $ G_{\text{RL}} $) in the differential conductance matrix at zero temperature is~\cite{datta1995electronic,anantram1996current, menard2020conductancematrix,anselmetti2019endtoend}
	\begin{equation}\label{eq:condmat}
		\hat{G}=\mqty*(G_{\text{LL}} & G_{\text{LR}}\\ G_{\text{RL}} & G_{\text{RR}})=\mqty*(\pdv{I_L}{V_L} & -\pdv{I_L}{V_R} \\ -\pdv{I_R}{V_L} & \pdv{I_R}{V_R}),
	\end{equation}
	where $ I_{\text{L}} $ ($ I_{R} $) is the current entering the left (right) normal lead from the scattering region, and $ V_{\text{L}} $ ($ V_{\text{R}} $) is the bias applied on the left (right) normal lead. Note that different sign conventions may be adopted~\cite{rosdahl2018andreev} on the nonlocal conductance, though it does not affect any real physical conclusion as long as we stick to one convention consistently. In Blonder-Tinkham-Klapwijk formalism,~\cite{blonder1982transition} the local conductance is
	\begin{equation}\label{eq:Glocal}
		G_{ii}=\frac{e^2}{h}\qty(N_i-T_{ii}^{ee}+T_{ii}^{eh}),
	\end{equation}
	and the nonlocal conductance is
	\begin{equation}\label{eq:Gnonlocal}
		G_{ij}=\frac{e^2}{h}\qty(T_{ij}^{ee}-T_{ij}^{eh}), \qquad i\neq j,
	\end{equation}
	where $ N=2 $ is the number of the electron mode in this single channel model, and the transmission $ T_{ij}^{\alpha\beta}=\tr(\qty[S_{ij}^{\alpha\beta}]^\dagger S_{ij}^{\alpha\beta}) $. {[$ \tr(..) $ for the trace]}

	The local conductance combined with the nonlocal conductance may be insufficient to distinguish the MBS from ABS in {realistic cases}, which are dominated by disorder effect. Therefore, we additionally calculate the thermal conductance which is defined as~\cite{akhmerov2011quantized,beenakker2015randommatrix}
	\begin{equation}\label{eq:Gth}
		\kappa=\kappa_0(T_{ij}^{ee}+T_{ij}^{eh}), \qquad i\neq j,
	\end{equation}
	where the quantized peak $ \kappa_0=\pi^2k_B^2\tau/6h $ at temperature $ \tau $ happens at TQPT~\cite{senthil1999spin,senthil2000quasiparticle,evers2008anderson} {($ h $ for the Planck constant and $ k_B $ for the Boltzmann constant)}.
	
	To identify the TI in the finite-length nanowire, we refer to the topological visibility (TV)~\cite{dassarma2016how,fulga2011scattering} defined as $ \det(S_{\text{LL}}) $ and $ \det(S_{\text{RR}}) $ with a positive value (+1, ideally) indicating trivial regime and a negative value (-1, ideally) indicating topological regime. Therefore, the TQPT is indicated by the zero crossings of the TV when the unitarity of $ \det(S_{\text{LL}}) $ or $ \det(S_{\text{RR}}) $ breaks. We additionally impose a phenomenological dissipation term $ i\Gamma $ with $ \Gamma=10^{-3} $ meV~\cite{liu2017role,dassarma2016how,liu2017phenomenology} in the Hamiltonian~\eqref{eq:H} to circumvent potential numerical singularities from local bound states in the middle of the wire. Such a small dissipation is invariably present in the experiments. We also set the height of barrier at the NS junction interface to be 0 when calculating the TV.

	\subsection{Quantum dots and inhomogeneous potential}
	
	The minimal effective model in Hamiltonian~\eqref{eq:bdg-pristine} shows a pristine nanowire example resulting in the good ZBCP. However, ZBCPs in nanowires used in present experiments are potentially more consistent with the so-called ``bad" and ``ugly" ZBCP.~\cite{pan2020physical} We first start with the bad ZBCP and will discuss the ugly ZBCP in the next subsection. 
	
	There are two main sources that can induce the bad ZBCP: The first source is when the lead is attached to the end of the SM, it is unavoidable that the contact point can create an unexpected quantum dot due to the mismatch of Fermi energy between the normal lead and the SM by creating a Schottky barrier.~\cite{liu2017andreev} Therefore, unintentional quantum dots at the interface are quite ubiquitous in experiments. In theory, the quantum dot can be effectively described by a potential barrier in the SM which is also uncovered by the SC. Therefore, the quantum dot plays a role in $ H_{\text{V}}=V(x) $ in the Hamiltonian~\eqref{eq:H}. Without any loss of generality,~\cite{liu2017andreev,moore2018quantized} we can parameterize the quantum dot by the barrier height $ V_{\text{D}} $ and the dot length $ l $ in the form of Gaussian potential, i.e., 
	\begin{equation}\label{eq:qd}
		H_{\text{V}}=V(x)=V_{\text{D}} \exp(-\frac{x^2}{l^2})\theta(l-x)
	\end{equation}
	and 
	\begin{equation}\label{eq:Delta0}
		\Delta_0(x)=\Delta_0 \theta(l-x),
	\end{equation}
	where $ \theta(x) $ is the Heaviside step function to account for the partially-covered SM.
	
	The second source of bad ZBCP is an inhomogeneous and smooth background potential,~\cite{kells2012nearzeroenergy,rainis2013realistic,stanescu2014nonlocality,stanescu2019robust,moore2018quantized} which may arise from the effect of gate voltage tuning the system or the charged impurities in the environment. In theory, the inhomogeneous potential also acts on the potential of SM as appearing in $ H_{\text{V}} $. We model the inhomogeneous potential using Gaussian potential~\cite{stanescu2019robust} again with the maximal peak $ V_{\text{max}} $ and the linewidth $ \sigma $, i.e.,
	\begin{equation}\label{eq:inhom}
		H_{\text{V}}=V(x)=V_{\text{max}} \exp(-\frac{x^2}{2\sigma^2}).
	\end{equation}
	We emphasize that both sources of bad ZBCP-- quantum dots and inhomogeneous potential--- create an effective spatially varying chemical potential $ \mu-V(x) $ in the BdG Hamiltonian~\eqref{eq:bdg-pristine}. They only differ in their physical mechanisms. Therefore, they both induce bad ZBCP, mimicking MZM, in a similar manner. 
	
	\subsection{onsite disorder}
	Unlike the quantum dot and inhomogeneous potential, which both induce the bad ZBCP, the onsite random disorder has threefold effects on the ZBCP: (1) weak disorder can leave the topological regime unchanged, preserving the good ZBCP; (2) strong disorder itself can induce trivial ugly ZBCP. The ugly ZBCP can mimic the good ZBCP in the local conductance measurement; (3) very strong disorder will simply destroy all SC in nanowire and leave the system in the Anderson localized phase.~\cite{anderson1958absence} The onsite disorder may arise in the chemical potential, parent SC gap, and effective $ g $ factor. In this paper, we will mainly discuss the effects of the disorder in the chemical potential, which is the leading disorder in SM nanowire.~\cite{yu2021nonmajorana,chen2019ubiquitous}
	
	The disorder in the chemical potential is added to $ H_{\text{V}} $ in Hamiltonian~\eqref{eq:H}, which is a phenomenological description of all kinds of onsite impurities in the SM.~\cite{liu2012zerobias,bagrets2012class,pikulin2012zerovoltage,brouwer2011topological,sau2013density,lin2012zerobias,neven2013quasiclassical,sau2013bound,degottardi2013majorana,adagideli2014effects,roy2013topologically,hui2015bulk} We denote the disorder in the chemical potential by $ H_{\text{V}}=V_{\text{imp}}(x) $. The white noise $ V_{\text{imp}}(x) $ is an onsite random potential that obeys the uncorrelated Gaussian distribution with the mean value of zero and standard deviation of $ \sigma_\mu $. Namely, $ \expval{V(x)V(x')}=\sigma_\mu^2\delta(x-x') $. We clarify that all the following results are calculated for just one specific configuration of randomness without the average over disorder as appropriate for a specific sample. The weak disorder preserves the good ZBCP; but strong disorder creates the ugly ZBCP--- the trivial ZBCP which emerges below the TQPT while the topological ZBCP still persists above the TQPT. For a very strong disorder, the disorder eventually destroys the topological ZBCP by creating a very large coherence length since the SC gap may collapse. {In such a strong disorder situation, the topological invariant, even for longer systems, tends to be trivial.} The strong disorder case is more appropriately studied by considering the system to be a random matrix.~\cite{pan2020generic}

	\section{results}\label{sec:results}
	In this section, we first show representative electrical conductance, thermal conductance, and TV of the good, bad, and ugly ZBCP in the three-terminal long nanowire in Secs.~\ref{sec:results-good},~\ref{sec:results-bad}, and ~\ref{sec:results-ugly}. In Sec.~\ref{sec:results-qdmuVar}, we add disorder to a bad ZBCP case with the quantum dot for a more realistic simulation. We also present results for the short wire in Sec.~\ref{sec:results-short} for the pristine wire and the nanowire in the presence of intermediate disorder with and without the parent SC gap collapse. Other examples of the bad and ugly ZBCP are presented in the appendix. We provide most of our results in the appendix to help with the readability of the text and not because the results in the appendix are less important.
	\subsection{Pristine nanowire}\label{sec:results-good}	
\begin{figure*}[htbp]
		\centering
		\includegraphics[width=6.8in]{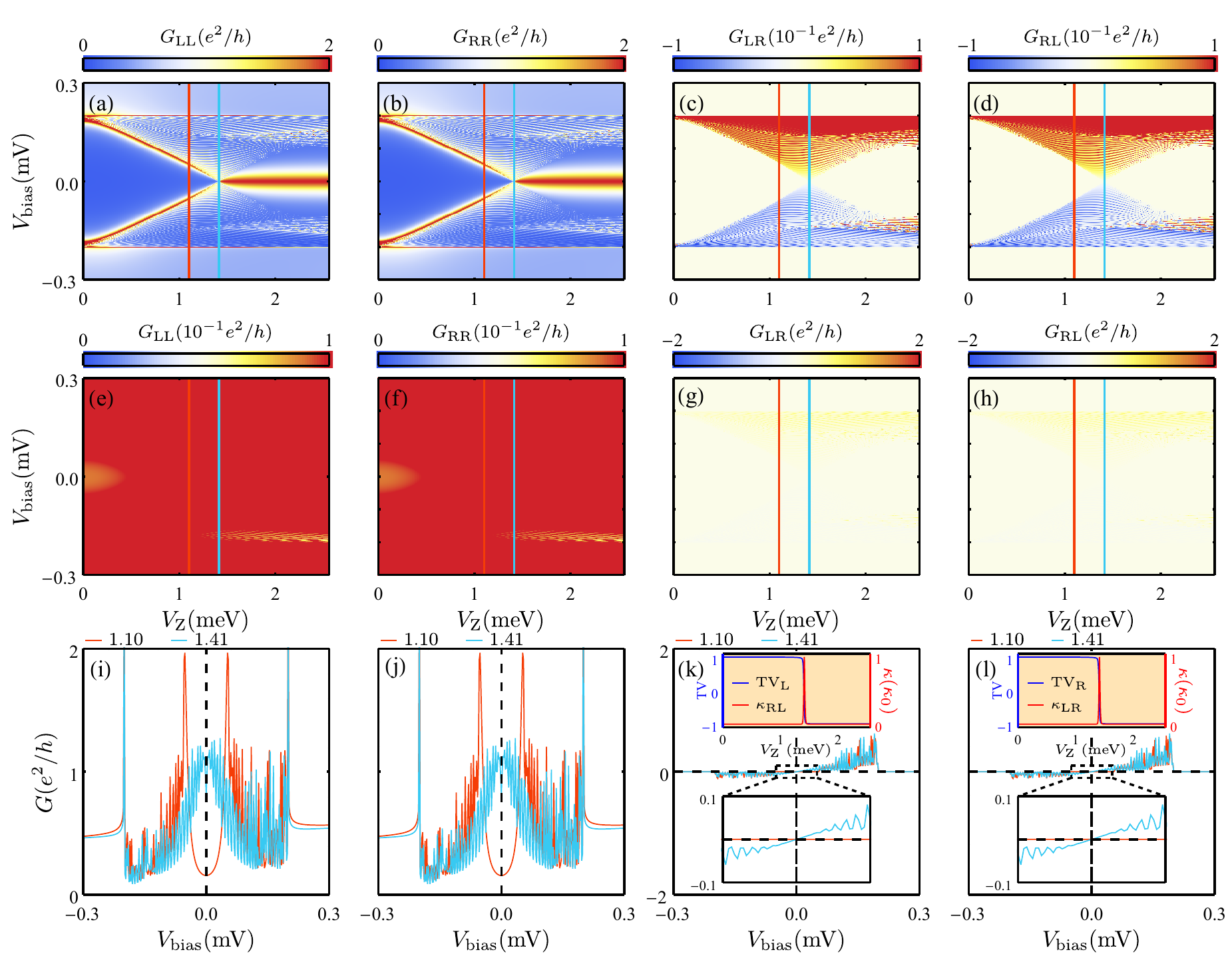}
		\caption{The good ZBCP in a pristine nanowire. (a)-(d) show the local and nonlocal conductance in the ``intrinsic'' color scale. (e)-(h) show the same conductance in the {opposite} color scale. (i)-(j)
			show line cuts of the conductance as a function of bias at $ V_{\text{Z}}=1.1 $ meV and $ 1.41 $ meV. The parameters are: chemical potential $ \mu=1 $ meV, parent SC gap $ \Delta_0=0.2 $ meV, SC-SM coupling strength $ \gamma=1 $ meV, SOC $ \alpha=0.5 $ eV\AA, the height of barrier {is} 5 meV, and wire length $ L=3\mu$m. The corresponding TV from the left (right) and thermal conductance $ \kappa_{RL} $ ($ \kappa_{LR} $) are shown in the inset of (k) [(l)].}
		\label{fig:good}
	\end{figure*}

	We start with the good ZBCP in a pristine nanowire as presented in Fig.~\ref{fig:good} to show how this proposal works since the good ZBCP provides a milestone of what the real Majorana should ideally look like in the three-terminal device in future experiments. We present the local conductance ($ G_{\text{LL}} $, $ G_{\text{RR}} $) and nonlocal conductance ($ G_{\text{LR}} $, $ G_{\text{RL}} $), respectively, from the first column to the last column along with the line cut of the conductance at different $ V_{\text{Z}} $. 

	The conductance in the first row is plotted in their ``intrinsic" (i.e., defined by the maximum and minimum values in the plot) color scales---the color bar ranges from $ 0 $ to $ 2e^2/h $ for the local conductance [Figs.~\ref{fig:good}(a) and (b)] and the color bar ranges from $ -0.1 e^2/h  $ to $ 0.1 e^2/h  $ for the nonlocal conductance [Figs.~\ref{fig:good}(c) and (d)]. The conductances $G_{\text{LL}}$ and $ G_{\text{RR}} $ plotted in Figs.~\ref{fig:good}(a) and~\ref{fig:good}(b) show the appearance of a quantized ZBCP at the TQPT (light blue line).~\cite{sau2010generic} The nonlocal conductance ($ G_{\text{LR}} $, $ G_{\text{RL}} $) shown in Figs.~\ref{fig:good}(c) and~\ref{fig:good}(d) provides a more direct measure of gap closure as opposed to the local conductance $ G_{ii} $. The {gap closure} features seen in $ G_{ii} $ have been identified previously to be associated with ABSs localized at each end of the nanowire. By contrast, the conductance $ G_{\text{LR}} $ seen in Fig.~\ref{fig:good} can only represent contributions from states that are delocalized across the nanowire. Thus, the transport gap closing and reopening seen in {Figs.}~\ref{fig:good}(c) and~\ref{fig:good}(d) provides direct evidence for a gap closure and reopening associated with TQPT. 
	
	In the second row, we present the same calculated conductance data in the opposite way: The local conductance is colored with the ``intrinsic'' scale for nonlocal conductance and vice versa. Although this may look absurd at the first glance--- the local conductance is almost saturated at the ZBCP and the nonlocal conductance is nearly invisible everywhere--- we still want to emphasize that this is what Majorana would look like in experiments of the three-terminal measurement. Even in the pristine wire, without any disorder or inhomogeneous potential, we almost see nothing in the nonlocal conductance because the nonlocal conductance [{Figs.}~\ref{fig:good}(g) and (h)] is several orders of magnitudes smaller than the local conductance. The only way we can observe any signature is to measure the conductance with very high precision ($ \sim10^{-2}e^2/h $). This is the premise of the three-terminal measurement. Recent experiments~\cite{menard2020conductancematrix,puglia2020closing} report such nonlocal conductance measurements with the precision up to $ 10^{-3}e^2/h $, in an attempt to use three-terminal measurements to distinguish the MBS from ABS. We emphasize that if the signal in any nonlocal three-terminal conductance measurement is comparable to (or even just an order of magnitude less than $ e^2/h $) the local conductance, then the only possible conclusion is that the measurement is dominated by noise and artifacts.  
	
	In the third row, we show the line cuts at specific magnetic field $ V_{\text{Z}} $ for the local and nonlocal conductance. The red line intersects a trivial regime and the light blue line is at the TQPT (we follow this convention throughout the paper). By presenting the four conductance plots ``redundantly'', we emphasize that the local conductance $ G_{\text{LL}} $ and $ G_{\text{RR}} $ are identical, and the same for nonlocal conductance $ G_{\text{LR}} $ and $ G_{\text{RL}} $. The inset at the critical point shows a linear dependence of the conductance on bias voltage.
	
	Measurement of thermal transport provides another indication of gap closure. As seen in the {insets} of {Fig}.~\ref{fig:good}(k) and (l), the thermal conductance shows a clear quantized peak~\cite{akhmerov2011quantized} indicating the topological quantum phase transition. While the thermal conductance might be intrinsically more challenging to measure, we note that the thermal conductance does not show any of the suppression near zero bias that the nonlocal electrical conductance shows. 
	
	The degree to which the pristine nanowire is topological can be characterized by the topological visibility, which is plotted in the {insets} of {Figs}.~\ref{fig:good}(k) and~\ref{fig:good}(l). The calculated TV for the pristine wire shows a clear transition from -1 to 1 at the the TQPT (light blue line), which occurs at the theoretically expected critical Zeeman field $ \sqrt{\gamma^2+\mu^2} $. Thus, in this case the light blue line represents the unambiguous characterization of when the nanowire is topological. The other panels of Fig.~\ref{fig:good} that show conductances that are in principle measurable in experiment show the expected characteristics of the {topological}, nontopological phase, as well as the TQPT. Unfortunately, TV, being a pure theoretical construct, cannot be measured in the laboratory, but it does enable the theory in ascertaining whether a particular experimental sample can in principle manifest topological properties.

	\subsection{Quantum dot and inhomogeneous potential}\label{sec:results-bad}
	Once we establish the protocol of three-terminal measurement in the pristine nanowire, we proceed to use this protocol in more realistic models. In this subsection, we consider the effect of a quantum dot and inhomogeneous potential on the nanowire, which can lead to a ZBCP~\cite{pan2020physical} similar to pristine topological nanowire seen in Figs.~\ref{fig:good}(a) and (b). This ZBCP may be good (i.e., topological) or bad (i.e., trivial) depending on the situation--- just looking at the ZBCP itself would not tell us whether the ZBCP is good or bad; one needs to know whether the ZBCP is below or above the TQPT to figure out its topological nature.

\begin{figure*}[htbp]
	\centering
	\includegraphics[width=6.8in]{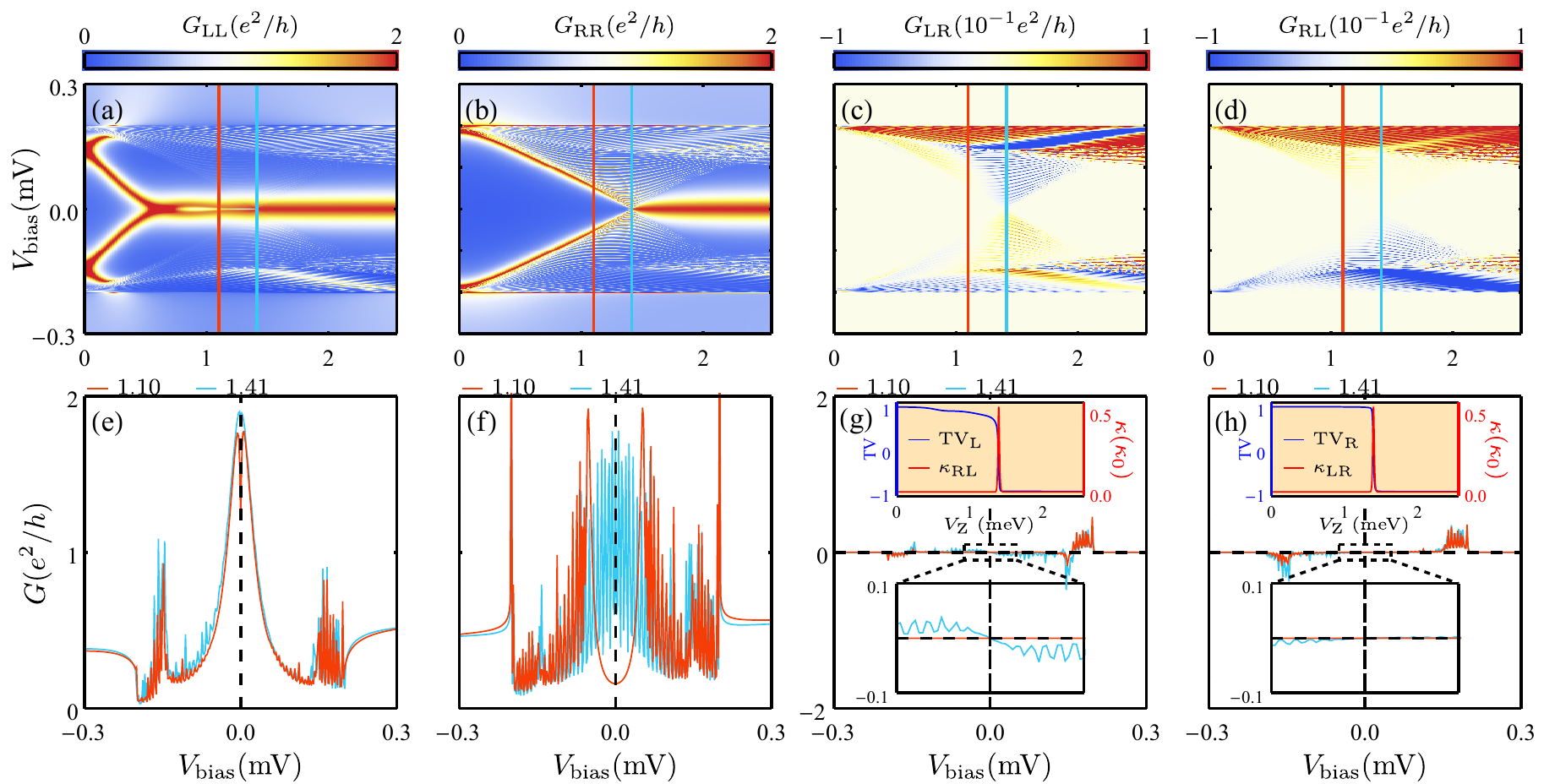}
	\caption{The bad ZBCP in a nanowire with the quantum dot on the left end. (a)-(d) show the local and nonlocal conductance in the ``intrinsic'' color scale. (e)-(h) are the corresponding line cuts of the conductance as a function of bias at $ V_{\text{Z}}=1.1 $ meV and $ 1.41 $ meV. The height of the quantum dot $ V_{\text{D}}=0.4 $ meV and the size $ l=0.15~\mu $m. The other parameters are the same as Fig.~\ref{fig:good}. The corresponding TV from the left (right) and thermal conductance $ \kappa_{RL} $ ($ \kappa_{LR} $) are shown in the inset of (g) [(h)].}
	\label{fig:qd}
\end{figure*}

We first add a quantum dot on the left end of the nanowire, and present the electrical conductance in Fig.~\ref{fig:qd}. In the local conductance Fig.~\ref{fig:qd}(a), we notice that the ZBCP starts to emerge below TQPT at $ V_{\text{Z}}\sim0.5 $ meV, which shows a quantized and robust ZBCP without oscillation.  In the nonlocal conductance as presented in Figs.~\ref{fig:qd}(c) and (d), we find that the bulk gap does not collapse until the advent of real TQPT at $ V_{\text{Z}}=1.41 $ meV. 

In the second panel, we show the line cuts of two $ V_{\text{Z}} $ with the red being in the trivial regime, and the light blue being real TQPT determined by the peak of thermal conductance. In Figs.~\ref{fig:qd}(g) and (h), the nonlocal conductance on the left end is much more prominent than that of the right end due to the presence of the quantum dot on the left end but not on the right end. Thus, the measurement of the ``bad" quantized ZBCP from a quantum dot can, in principle, be distinguished from a MBS by the lack of an associated gap closure in Fig.~\ref{fig:qd}(c) at the value of Zeeman field where the ZBCP starts in Fig.~\ref{fig:qd}(a). At the same time, Fig.~\ref{fig:qd}(c) shows that the presence of the quantum dot might make the gap closure harder to see compared to the pristine case. 

As in the pristine case, the thermal conductances in the {insets} in Figs.~\ref{fig:qd}(g) and~\ref{fig:qd}(h) show quantized peaks at exactly the TQPT. For reference, we plot the TV in the inset as well, which indicates the position of the TQPT by where the TV vanishes. The deviations from quantization for both the thermal quantization and TV are a result of the finite dissipation we use to stabilize our numerical calculations.

\begin{figure*}[htbp]
	\centering
	\includegraphics[width=6.8in]{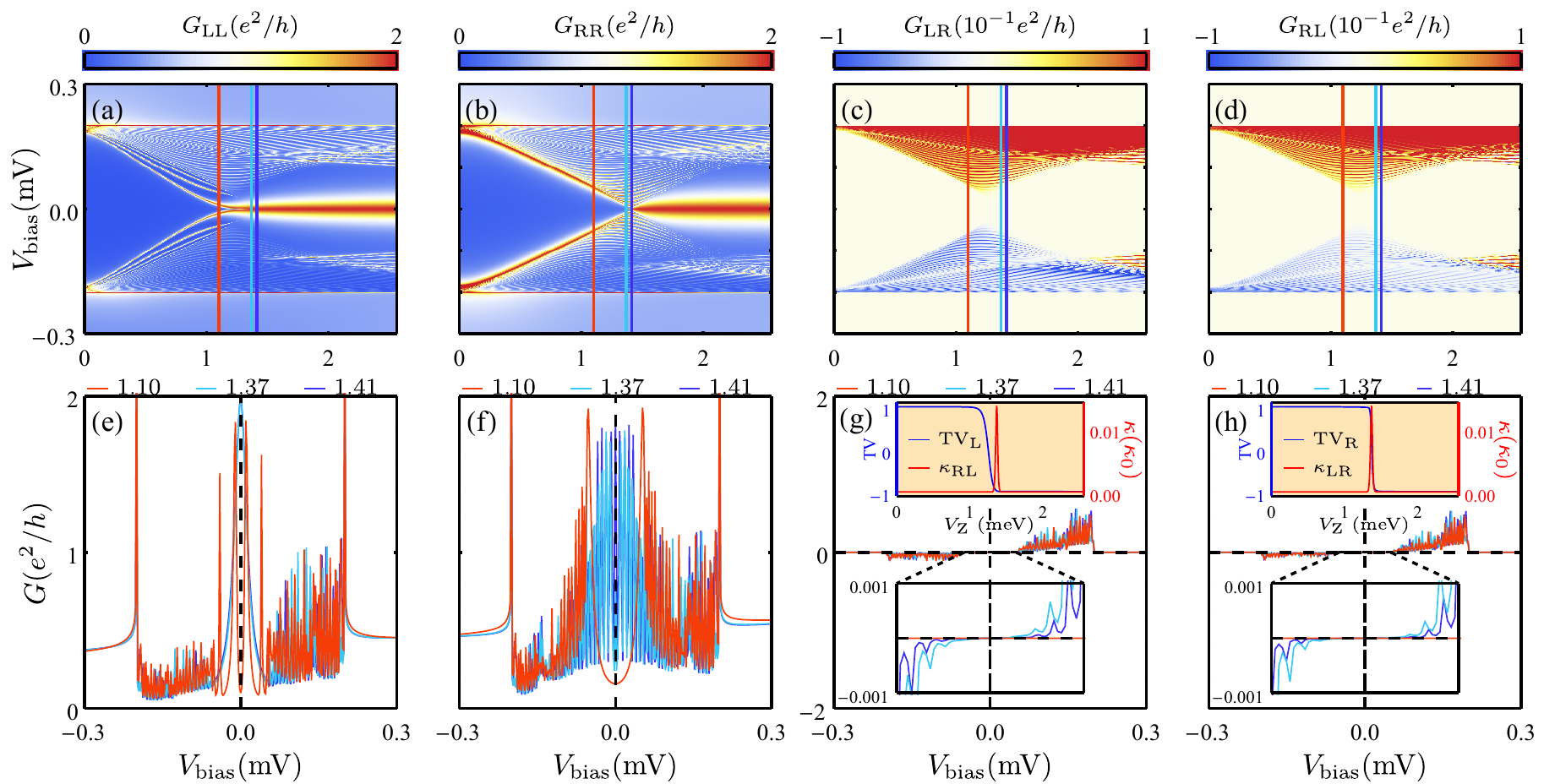}
	\caption{The bad ZBCP in a nanowire with the Gaussian inhomogeneous confinement on the left end. (a)-(d) show the local and nonlocal conductance in the ``intrinsic'' color scale. (e)-(h) are the corresponding line cuts of the conductance as a function of bias at $ V_{\text{Z}}=1.1 $ meV, 1.37 meV, and $ 1.41 $ meV. The height of the confining potential $V_{\text{max}}=1  $ meV and linewidth $ \sigma=0.45~\mu $m. The other parameters are the same as Fig.~\ref{fig:good}. The corresponding TV from the left (right) and thermal conductance $ \kappa_{RL} $ ($ \kappa_{LR} $) are shown in the inset of (g) [(h)].}
	\label{fig:inhom}
\end{figure*}

	The second source of the bad ZBCP is the inhomogeneous potential. We present the results of the inhomogeneous potential with one potential barrier on the left end in Fig.~\ref{fig:inhom}. The local conductance in Fig.~\ref{fig:inhom}(a) is similar to the quantum dot situation in Fig.~\ref{fig:qd}(a), where we also observe a quantized and robust ZBCP without any Majorana oscillation. However, in the nonlocal conductance in Figs.~\ref{fig:inhom}(c) and~\ref{fig:inhom}(d), we find that the {gap closure} and reopening feature is not salient--- the minimal gap does not completely close at the putative TQPT. Nevertheless, we can still distinguish the trivial zero-energy ABS from the MBS by the lack of associated gap closure at where ZBCP starts to emerge in Fig.~\ref{fig:inhom}(a).

	In the second panel, we additionally show the dark blue line of the {nominal} TQPT which is always fixed at $ V_{\text{Z}}=1.41 $ meV. Here, unlike the pristine wire and the quantum dot case, the value of the real TQPT determined from the peak of thermal conductance is smaller than that of the nominal TQPT, but the discrepancy is only slight (i.e., within the width of the thermal conductance peak and thus can be thought of as a finite size effect). The nonlocal conductance at the real TQPT [light blue in the {insets} of Figs.~\ref{fig:inhom}(g) and~\ref{fig:inhom}(h)] is also slightly larger than that at the nominal TQPT [large blue in the {insets} of Figs.~\ref{fig:inhom}(g) and (h)] near zero bias.

	Similar to the pristine wire and quantum dot, we again calculate the TV and the thermal conductance in the {insets} of {Figs.}~\ref{fig:inhom}(g) and~\ref{fig:inhom}(h). We find that the thermal conductance, though not quantized due to the finite dissipation, accurately indicates the position of TQPT where the TV flips the sign. 

	\subsection{Disorder in the chemical potential}\label{sec:results-ugly}
	
	Finally, we present the effect of disorder and show what the nonlocal conductance will look like in the presence of disorder. Since the weak disorder (typically $ \sigma_\mu/\mu\le1 $ in this configuration) preserves the good ZBCPs,~\cite{pan2020physical} we will skip our results for the nonlocal conductance in the weak disorder case, which is similar to the pristine case in Fig.~\ref{fig:good}. Weak disorder poses no problem to the detection of MZMs.

	\begin{figure*}[htbp]
	\centering
	\includegraphics[width=6.8in]{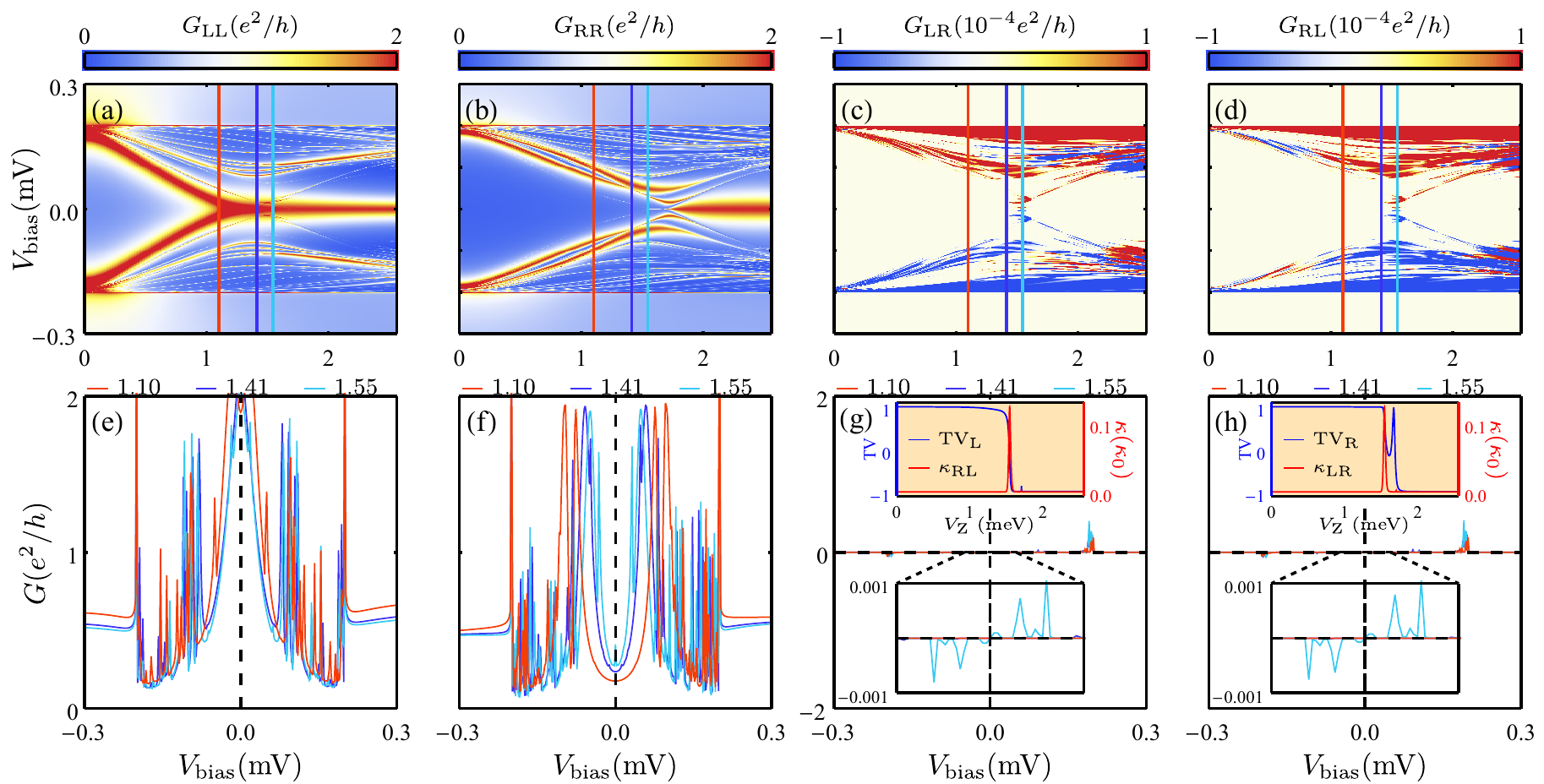}
	\caption{The ugly ZBCP in a nanowire in the presence of intermediate disorder with $ \sigma_\mu/\mu=1.5 $. (a)-(d) show the local and nonlocal conductance in the ``intrinsic'' color scale. (e)-(h) are the corresponding line cuts of the conductance as a function of bias at $ V_{\text{Z}}=1.1 $ meV, $ 1.41 $ meV, $ 1.55 $ meV. The other parameters are the same as Fig.~\ref{fig:good}. The corresponding TV from the left (right) and thermal conductance $ \kappa_{RL} $ ($ \kappa_{LR} $) are shown in the inset of (g) [(h)].}
	\label{fig:muVar1p5}
\end{figure*}

	We first add the intermediate disorder in the chemical potential with $ \sigma_\mu/\mu=1.5 $ as shown in Fig.~\ref{fig:muVar1p5}.  In the first row, we find that the trivial ZBCP emerges below the TQPT in $ G_{\text{LL}} $ (see Fig.~\ref{fig:muVar1p5}(a)) and above TQPT in {$ G_{\text{RR}} $} on the right [Fig.~\ref{fig:muVar1p5}(b)]. The left-right symmetry breaking in this case is accidental. Thus the local conductance resembles {Figs.}~\ref{fig:qd}(a) and~\ref{fig:qd}(b). In the nonlocal conductance, it looks like a signal of {gap closure} and gap reopening, though the signal is actually very small (we use a much smaller scale of colorbar here). Similar to the case of the pristine and bad ZBCPs the nonlocal conductance shown in Figs.~\ref{fig:muVar1p5}(c),~\ref{fig:muVar1p5}(d),~\ref{fig:muVar1p5}(g), and~\ref{fig:muVar1p5}(h) still shows a gap closure at the TQPT, though the signal is reduced by three orders of magnitudes compared to the already difficult to measure pristine case.
	
	In the second row of Fig.~\ref{fig:muVar1p5}, we present the line cuts of the color plots from the first row at Zeeman values representing the trivial regime (red), the nominal TQPT (dark blue), and the real TQPT determined by the peak of thermal conductance (light blue). The nonlocal conductance near zero bias also manifests an antisymmetric shape in the {insets} of Figs.~\ref{fig:muVar1p5}(g) and~\ref{fig:muVar1p5}(h). 
	
	The insets of Figs.~\ref{fig:muVar1p5}(g) and~\ref{fig:muVar1p5}(h) show the thermal conductance (plotted in red). The thermal conductance in this case shows a peak at the TQPT as in the pristine case. The TV ({plotted} in blue in the insets in Figs.~\ref{fig:muVar1p5}(g) and (h)) shows a well-defined topological regime, though there are few small spikes due to the disorder in the TV. The topological regime here is also narrower than that in the pristine wire in the sense that the trivial regime extends above the nominal TQPT ($ \sqrt{\mu^2+\gamma^2}=1.41 $ meV). Of course, the nominal TQPT does not indicate a real phase transition here. {Nevertheless}, we still show this line cut [dark blue lines in {Figs.}~\ref{fig:muVar1p5}(g) and (h)] to emphasize that the real disorder-renormalized TQPT is more and more deviated from the nominal TQPT (1.41 meV) in the pristine limit as disorder increases.

	\begin{figure*}[htbp]
		\centering
		\includegraphics[width=6.8in]{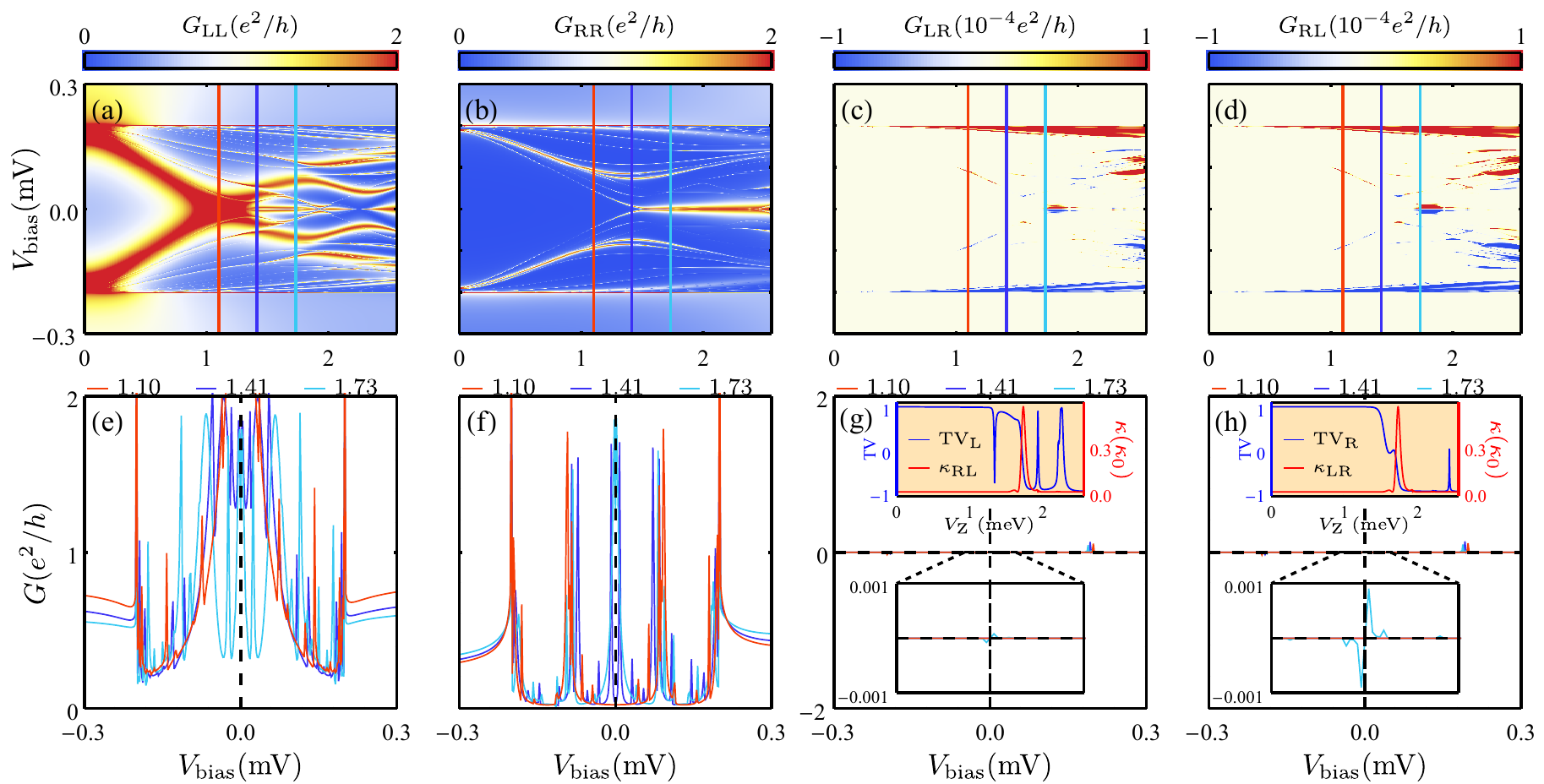}
		\caption{The ugly ZBCP in a nanowire in the presence of strong disorder with $ \sigma_\mu/\mu=3 $. (a)-(d) show the local and nonlocal conductance in the ``intrinsic'' color scale. (e)-(h) are the corresponding line cuts of the conductance as a function of bias at $ V_{\text{Z}}=1.1 $ meV, $ 1.41 $ meV, $ 1.73 $ meV. The other parameters are the same as Fig.~\ref{fig:good}. The corresponding TV from the left (right) and thermal conductance $ \kappa_{RL} $ ($ \kappa_{LR} $) are shown in the inset of (g) [(h)].}
		\label{fig:muVar3}
	\end{figure*}

	Next, we increase the disorder to $ \sigma_\mu/\mu=3 $ as shown in Fig.~\ref{fig:muVar3}. In the first row of Fig.~\ref{fig:muVar3}, the trivial ZBCP emerges while the topological ZBCP is completely destroyed on the left end due to the large disorder. In the nonlocal conductance, the signature of gap closing is very faint and the signature of gap reopening is absent even if we use a color scale with a very small range of conductance. 
	
	In the second row of Fig.~\ref{fig:muVar3}, we present the line cuts of the color plots of the nonlocal conductance from the first row and find vanishingly small signal almost everywhere. The upper insets of Figs.~\ref{fig:muVar3}(g) and~\ref{fig:muVar3}(h) show the thermal conductance (red line). The thermal conductance has the peak with a height smaller than the quantized value due to the finite dissipation. The TV (blue line) in the insets of Figs.~\ref{fig:muVar3}(g) and~\ref{fig:muVar3}(h) oscillates abruptly as the magnetic field increases. 

	\begin{figure*}[htbp]
	\centering
	\includegraphics[width=6.8in]{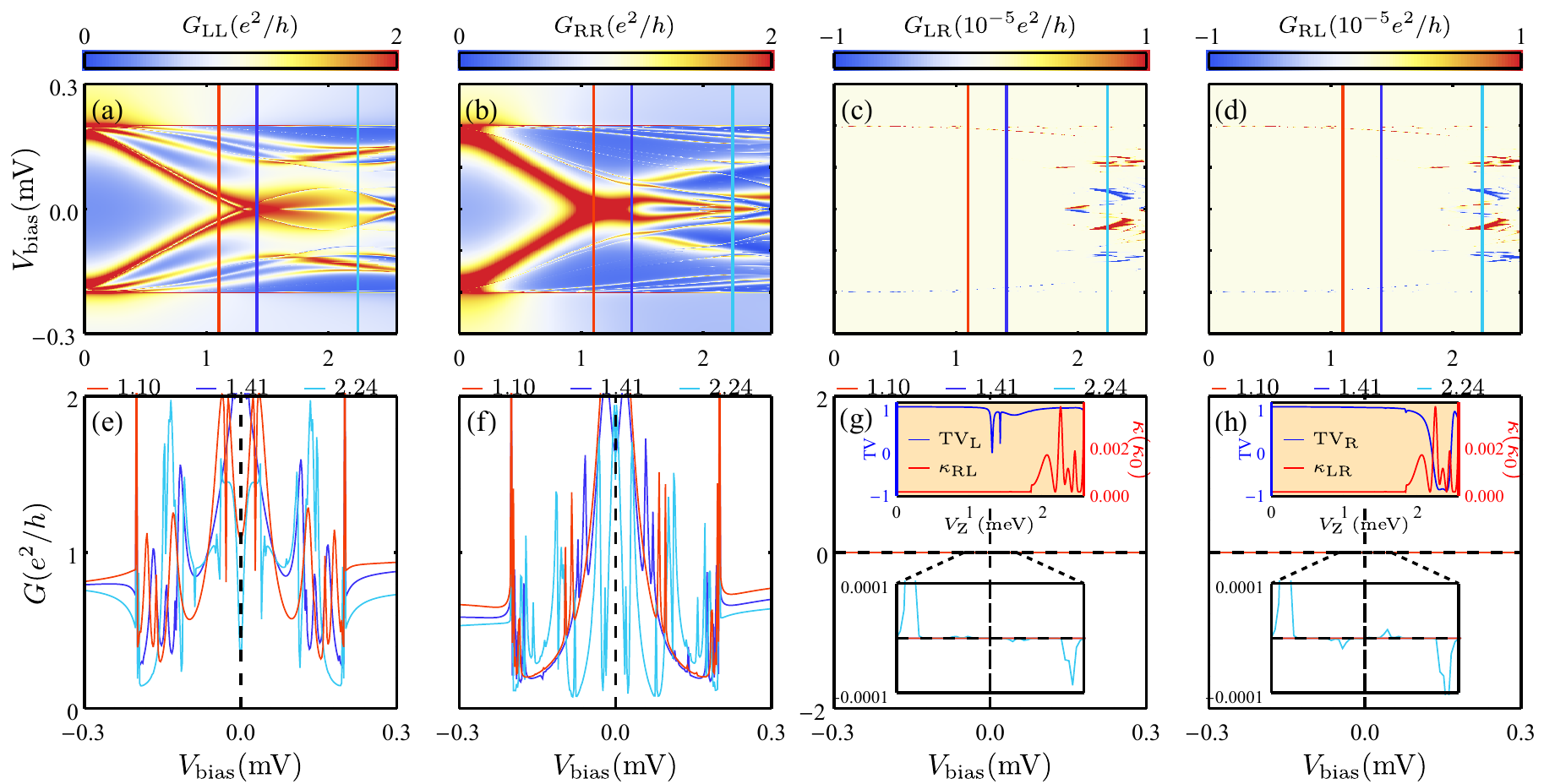}
	\caption{The ugly ZBCP in a nanowire in the presence of very strong disorder with $ \sigma_\mu/\mu=5 $. (a)-(d) show the local and nonlocal conductance in the ``intrinsic'' color scale. (e)-(h) are the corresponding line cuts of the conductance as a function of bias at $ V_{\text{Z}}=1.1 $ meV, $ 1.41 $ meV, $ 2.24 $ meV. The other parameters are the same as Fig.~\ref{fig:good}. The corresponding TV from the left (right) and thermal conductance $ \kappa_{RL} $ ($ \kappa_{LR} $) are shown in the inset of (g) [(h)].}
	\label{fig:muVar5}
\end{figure*}
	Finally, we use a very large disorder $  \sigma_\mu/\mu=5 $ and present the electrical and thermal conductance in Fig.~\ref{fig:muVar5}. In the local conductance in the first row, the ZBCP are all trivially induced by disorder; the ZBCP above the nominal TQPT is completely absent. The nonlocal conductances in Figs.~\ref{fig:muVar5}(c) and (d) are featureless: Almost everywhere is measured a tiny conductance within $ 10^{-5} e^2/h$.
	
	In the second row of Fig.~\ref{fig:muVar5}, we present the line cuts of the color plots of the nonlocal conductance from the first row, where, again, the conductance is almost zero. The upper insets of Figs.~\ref{fig:muVar5}(g) and~\ref{fig:muVar5}(h) show the thermal conductance (red line). The predicted thermal conductance in {this} strong disorder case goes from essentially vanishing at low Zeeman field to an irregular structure with many peaks, suggesting suppression of the SC transmission gap with magnetic field. Thus, the thermal conductance in this case does not appear to indicate any topological phase. This is consistent with the TV [blue line in the insets of Figs.~\ref{fig:muVar5}(g) and (h)], which does not flip to -1.
		
	In fact, the ugly ZBCP in Fig.~\ref{fig:muVar1p5} continuously transmutes to Anderson localization in Fig.~\ref{fig:muVar5} when disorder increases from the intermediate level to the very strong level. To show this continuous development, we additionally present other examples of intermediate and strong disorder with $ \sigma_{\mu}/\mu$ ranging from 2 up to 4.5 in Appendix~\ref{app:A} for a direct comparison.
	
	Before ending this subsection, we note that the scale of the disorder $ \sigma_\mu/\mu $ where disorder effects become substantial is significantly higher compared to our earlier work~\cite{pan2020generic} on the subject. This is largely due to our using a larger tunnel coupling strength $ \gamma $. In general the effect of chemical potential disorder on the system depends on other parameters, e.g., the tunnel coupling (which is usually unknown in experiments) and the lattice constant of discretization (which is purely artificial). The tunnel coupling strength $ \gamma $ sets the overall strength of the conductance and also determines the effective SC gap at zero energy;~\cite{stanescu2010proximity} therefore, it changes the effective coherence length. Since we use a larger tunnel coupling strength $ \gamma=1 $ meV other than the previous $ 0.2 $ meV, the critical value increases due to the larger SC gap protection. Therefore, one should not take the value of disorder $ \sigma_{\mu}/\mu=1.5 $ and conclude that these results are in the strong disorder limit. In fact, in our simulations, since we choose the disorder and keep other parameters fixed throughout, we can directly compare them and the relative results are sensible. However, the matter of how to directly map this disorder in our theory to experimental disorder values in real samples is unknown at this stage. Each sample will have its own weak to intermediate to strong disorder regime depending on the unknown details of various sample parameters such as tunnel coupling, SC gap, etc.
	
	\subsection{Quantum dot with disorder in the chemical potential}\label{sec:results-qdmuVar}
\begin{figure*}[htbp]
	\centering
	\includegraphics[width=6.8in]{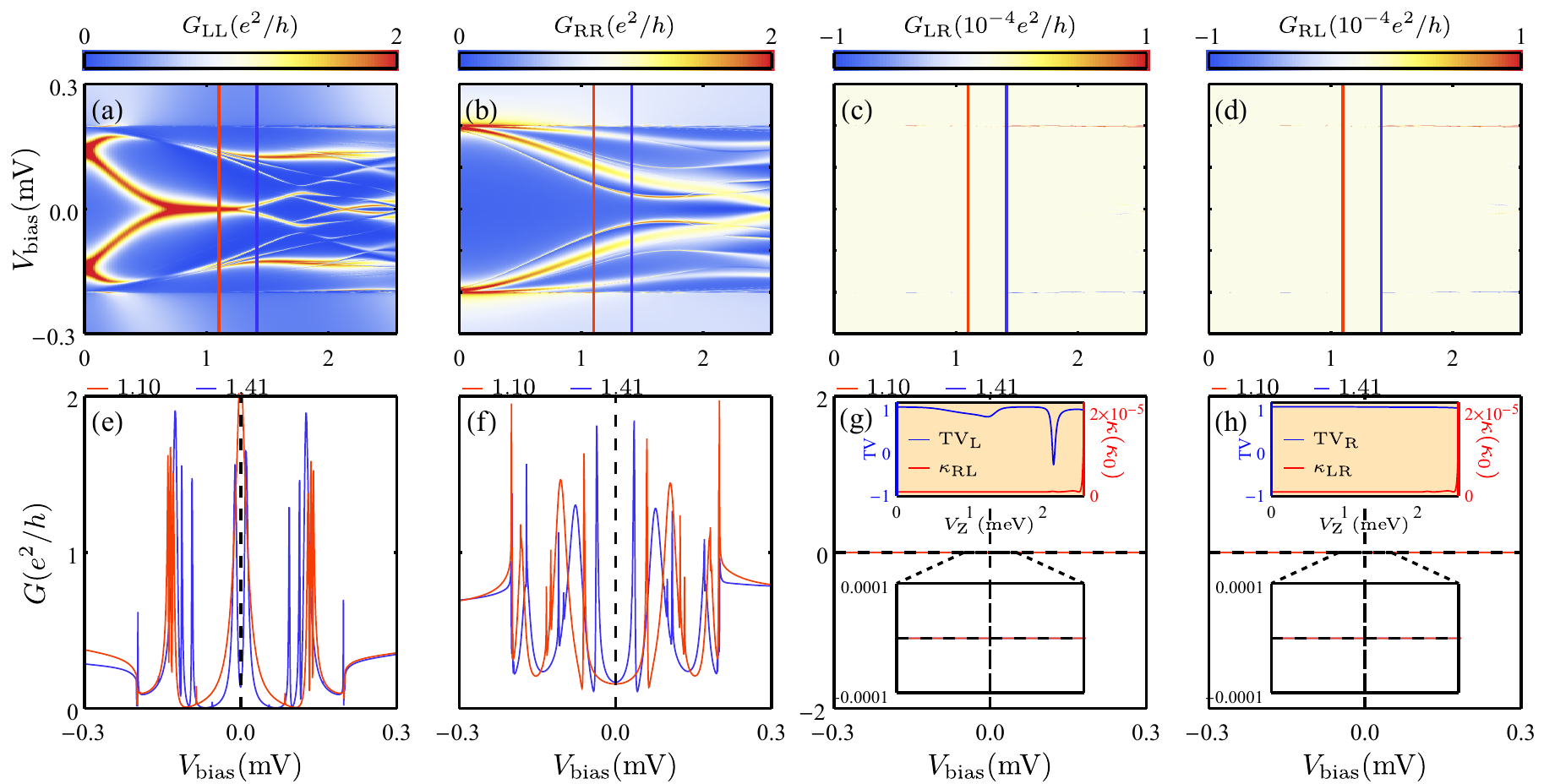}
	\caption{The wire in the presence of the quantum dot and very strong disorder with $ \sigma_\mu/\mu=5 $. (a)-(d) show the local and nonlocal conductance in the ``intrinsic'' color scale. (e)-(h) are the corresponding line cuts of the conductance as a function of bias at $ V_{\text{Z}}=1.1 $ meV, and $ 1.41 $ meV. The quantum dot is the same as Fig.~\ref{fig:qd} with $ V_{\text{D}}=0.4 $ meV and $ l=0.15~\mu $m. The other parameters are the same as Fig.~\ref{fig:good}. The corresponding TV from the left (right) and thermal conductance $ \kappa_{RL} $ ($ \kappa_{LR} $) are shown in the inset of (g) [(h)]. The nonlocal conductance is very small in this situation.}
	\label{fig:qdmuVar5}
\end{figure*}
	Finally, we consider a more complicated, and more realistic, situation where the quantum dot and disorder coexist. This is like mixing the bad ZBCP with ugly ZBCP. Although we deliberately isolate them and attribute them to different mechanisms, e.g., quantum dots or disorder, this is just for the better instructive purpose and there is no reason that only one of these mechanisms would happen one at a time. Therefore, it is natural to consider their combined effects on the nanowire. For example, we impose the disorder on the nanowire in the presence of a fixed quantum dot in Fig.~\ref{fig:qd}. The size of the quantum dot $ l=0.15~\mu $m and the height $ V_{\text{D}}=0.4 $ meV as defined in Eqs.~\eqref{eq:qd} and~\eqref{eq:Delta0}; the disorder in the chemical potential exists only in the region $ [l,L] $ of the nanowire, leaving the quantum dot intact. 
	
	By tuning the magnitude of disorder, we find that: If the disorder is weak, the bad ZBCP almost remains the same as if no disorder is present, which is expected. We present weak disorder results in Fig.~\ref{fig:qdmuVar0p5} in Appendix~\ref{app:A} and skip it in the main text because it is qualitatively the same as Fig.~\ref{fig:qd}. If the disorder is intermediate, e.g., $ \sigma_\mu/\mu=3 $ as shown in Fig.~\ref{fig:qdmuVar3} in Appendix~\ref{app:A}, we can still see the imprint of the quantum dot since the wire is still dominated by the quantum dot. 
	
	However, if the disorder is very large, for example $ \sigma_\mu/\mu=5 $ as shown in Fig.~\ref{fig:qdmuVar5}, apart from some ZBCP signatures of the quantum dot, the local conductance of the wire is qualitatively similar to the strongly disordered wire [see Figs.~\ref{fig:muVar5}(a) and (b)]. In $ G_{\text{LL}} $, the remnant effect of the quantum dot still shows up in the low-lying state near $ V_{\text{Z}}=0.5 $ meV while the topological regime is completely destroyed by the strong disorder. In Figs.~\ref{fig:qdmuVar5}(c) and~\ref{fig:qdmuVar5}(d), the nonlocal conductance is also very small everywhere, which is the same scenario as the strong disorder case discussed in Sec.~\ref{sec:results-ugly}. 
	
	In the second row, the line cuts of the color plots of the nonlocal conductance from the first row also show vanishingly small conductance. The thermal conductance [red line in the insets of {Figs}.~\ref{fig:qdmuVar5}(g) and~\ref{fig:qdmuVar5}(h)] remains vanishingly small, which is consistent with the TV. We find the TV [blue line in the insets of {Figs}.~\ref{fig:qdmuVar5}(g) and~\ref{fig:qdmuVar5}(h)] does not flip to -1, which again indicates that the wire continues to remain in the topologically trivial phase.

	\subsection{Short nanowires}\label{sec:results-short}
	\begin{figure*}[htbp]
		\centering
		\includegraphics[width=6.8in]{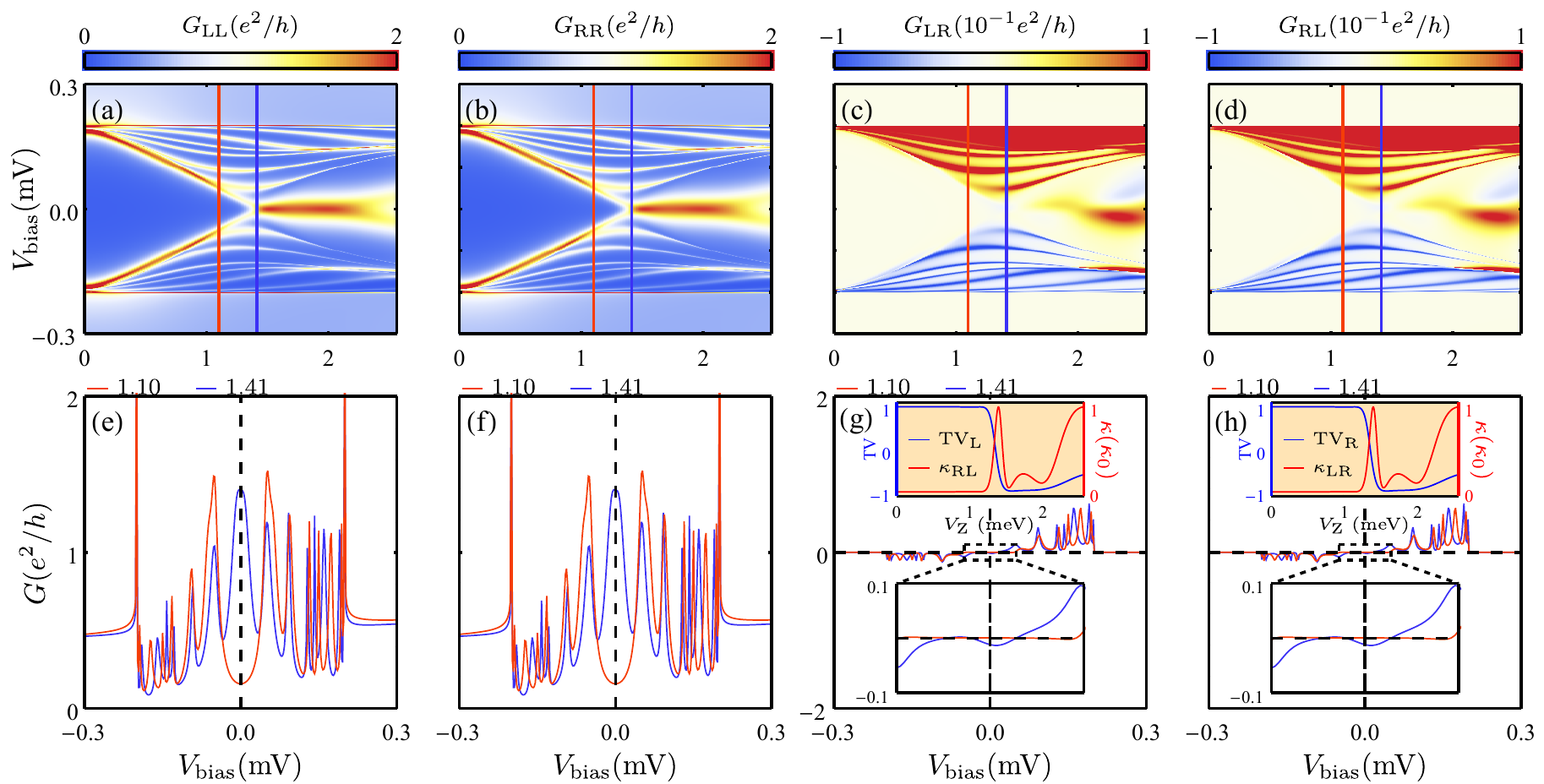}
		\caption{The short pristine nanowire ($ L=0.5~\mu $m). (a)-(d) show the local and nonlocal conductance in the ``intrinsic'' color scale. (e)-(h) show line cuts of the conductance as a function of bias at $ V_{\text{Z}}=1.1 $ meV and $ 1.41 $ meV. The other parameters are the same as Fig.~\ref{fig:good}. The corresponding TV from the left (right) and thermal conductance $ \kappa_{RL} $ ($ \kappa_{LR} $) are shown in the inset of (g) [(h)].}
		\label{fig:good0.5}
	\end{figure*}
	Previously, our results of the good, bad, and ugly ZBCP are all calculated under the long wire limit, where the magnitude of TV was close to unity so that wires could be precisely characterized as either topological or trivial. However, at the current stage of experiments, nanowires are of effective length up to $ 0.5\sim 1~\mu $m,~\cite{nichele2017scaling,zhang2018quantized} which is closer to the short wire limit (smaller than SC coherence length), even if some parameters that determine the coherence length in the nanowire are unknown in experiments, e.g., the SC-SM coupling strength and chemical potential. Therefore, we also present four representative short wire results in Fig.~\ref{fig:good0.5} to Fig.~\ref{fig:muVar3Sc} and show the other examples of short wires in Appendix~\ref{app:A}.
	
	In Fig.~\ref{fig:good0.5}, we use the pristine wire again in a shorter wire $ L=0.5~ \mu$m. The local conductance [Figs.~\ref{fig:good0.5}(a) and (b)] has a larger Majorana oscillation (which bifurcates at around $ V_{\text{Z}}\sim 2.5 $ meV) compared to the long wire situation in Fig.~\ref{fig:good}. In the nonlocal conductance [Figs.~\ref{fig:good0.5}(c) and (d)], we also see {gap closure} and reopening features but the minimal gap at the TQPT (see in the color plot) is not as small as that in the long wire limit.
	
	In the second row, we present the line cuts of the color plots of the nonlocal conductance from the first row, and find the nonlocal conductance shows significant nonlinearity at the TQPT around zero bias [light blue line in Figs.~\ref{fig:good0.5}(g) and~\ref{fig:good0.5}(h)], as opposed to the long wire limit in Fig.~\ref{fig:good}. Although the thermal conductance [red lines in the insets of Figs.~\ref{fig:good0.5}(g) and (h)] peaks at the quantized value at TQPT $ V_{\text{Z}}=1.41 $ meV, it grows again at a larger Zeeman field $ V_{\text{Z}}=2.5 $ meV as a result of the highly-overlapped Majorana. The TV [blue lines in the insets of Figs.~\ref{fig:good0.5}(g) and~\ref{fig:good0.5}(h)] also becomes less topological (more deviation from -1) at larger magnetic fields. This is because the coherence length generally increases as $ V_{\text{Z}} $ increases; therefore, the effective length of the wire becomes shorter at large magnetic fields.

	\begin{figure*}[htbp]
		\centering
		\includegraphics[width=6.8in]{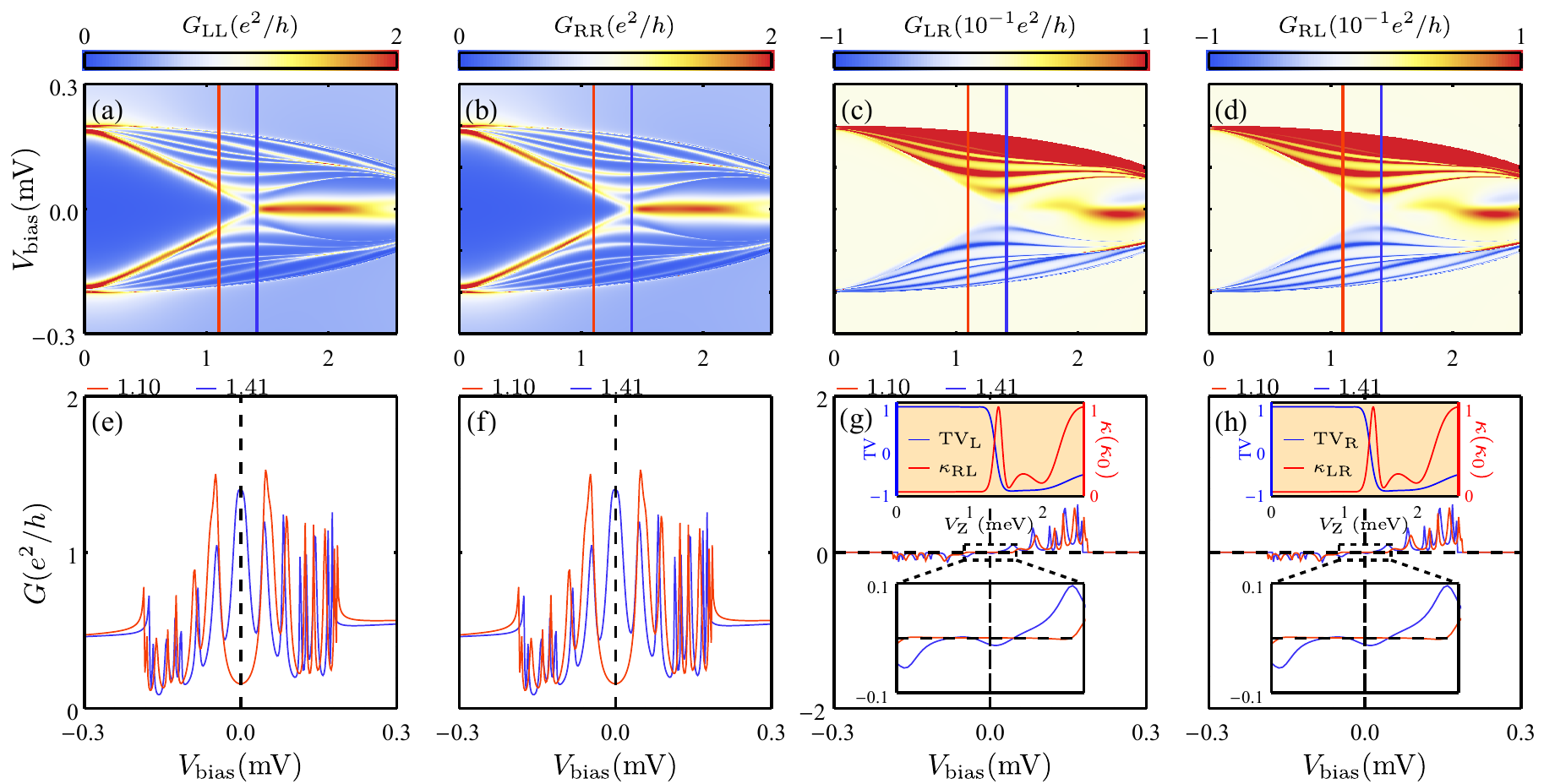}
		\caption{The short pristine nanowire ($ L=0.5~\mu $m). The SC bulk gap collapses as $ V_{\text{Z}} $ increases and closes completely at $ V_{\text{Z}}=3 $ meV (not shown here).  (a)-(d) show the local and nonlocal conductance in the ``intrinsic'' color scale. (e)-(h) show line cuts of the conductance as a function of bias at $ V_{\text{Z}}=1.1 $ meV and $ 1.41 $ meV. The other parameters are the same as Fig.~\ref{fig:good}. The corresponding TV from the left (right) and thermal conductance $ \kappa_{RL} $ ($ \kappa_{LR} $) are shown in the inset of (g) [(h)].}
		\label{fig:good0.5c}
	\end{figure*}	
		
	In experiments, the parent SC bulk gap often collapses as the magnetic field increases (most likely because of the penetration of the magnetic field into the parent Al). Therefore, we additionally introduce a phenomenological bulk gap collapse in the short wire as shown in Fig.~\ref{fig:good0.5c}. The corresponding modification to the Hamiltonian~\eqref{eq:bdg-pristine} is the change of the parent SC bulk gap, which closes in the form of $\Delta_0(V_{\text{Z}})=\Re(\sqrt{1-\qty(V_{\text{Z}}/V_{\text{Zc}})^2})\Delta_0 $. Here $ V_{\text{Zc}} $ is the Zeeman field that completely destroys the SC bulk gap. {We manually} set {it} to be $ V_{\text{Zc}}=3 $ meV. 
	
	In the local conductance in Figs.~\ref{fig:good0.5c}(a) and~\ref{fig:good0.5c}(b), we find the topological ZBCP is not affected by the collapse of parent SC gap as long as the parent SC gap does not vanish below TQPT. In the nonlocal conductance in Figs.~\ref{fig:good0.5c}(c) and~\ref{fig:good0.5c}(d), we find qualitatively similar results as in Fig.~\ref{fig:good0.5} with suppressed oscillations from overlapping Majoranas.
	
	In the second row, we present the line cuts (light blue lines) of the color plots of the nonlocal conductance from the first row and find that the conductance at zero bias is not altered by the collapse of the SC gap. However, the whole range of finite nonlocal conductance is suppressed as a result of a smaller parent SC gap. The thermal conductance [red lines in the upper {insets} of Figs.~\ref{fig:good0.5c}(g) and~\ref{fig:good0.5c}(h)] shows the exact same conductance as in Fig.~\ref{fig:good0.5} where the SC bulk gap does not collapse. The TV [blue lines in the upper {insets} of Figs.~\ref{fig:good0.5c}(g) and~\ref{fig:good0.5c}(h)] shows a peak around $ V_{\text{Z}}=1.41 $ meV and grows a second peak at $ V_{\text{Z}}=2.5 $ meV again. This implies that the effect of SC gap collapse is only quantitative--- it will not change topological {properties} of the nanowire as long as the SC gap does not vanish already below TQPT.
\begin{figure*}[htbp]
	\centering
	\includegraphics[width=6.8in]{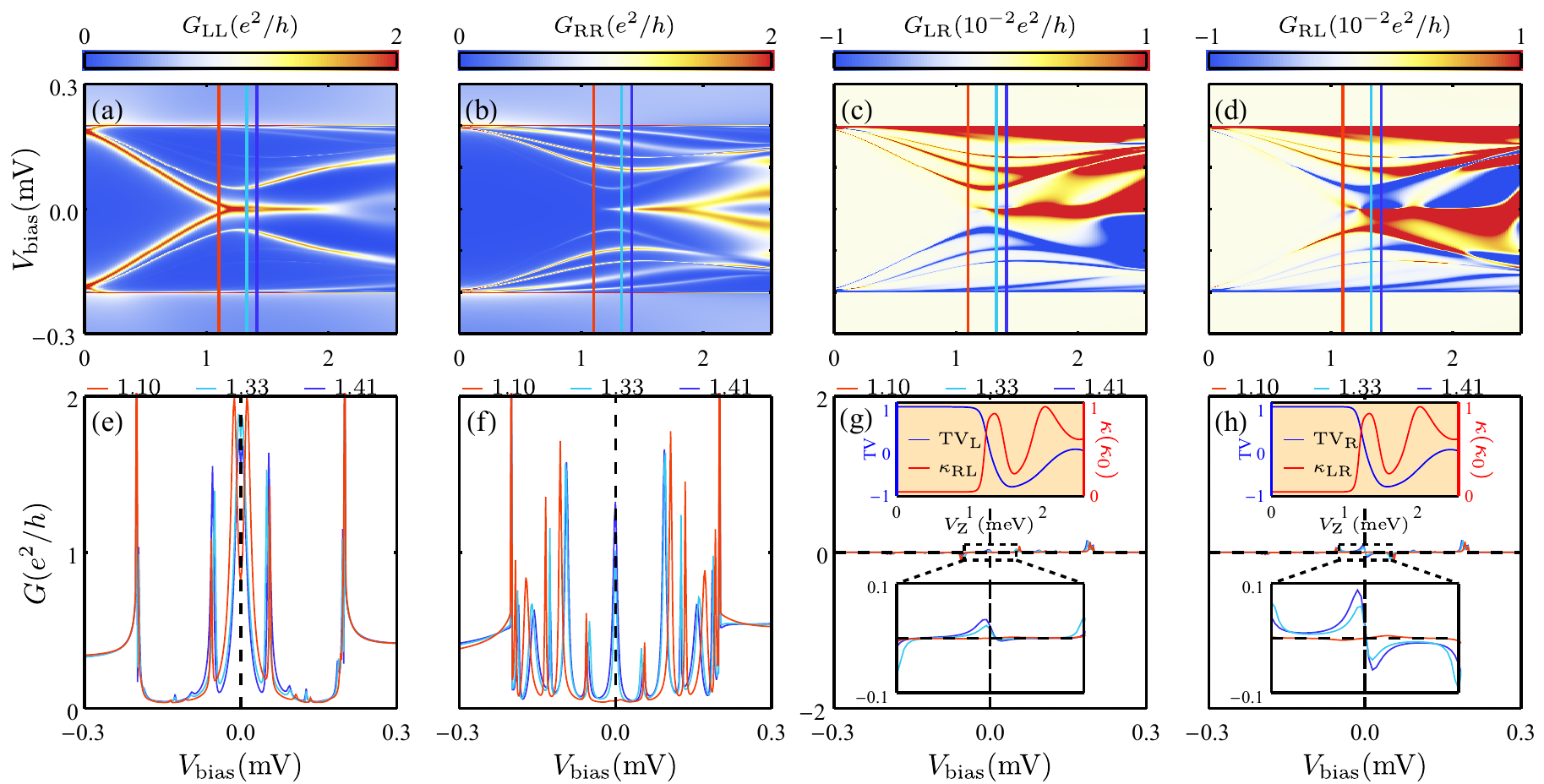}
	\caption{The short wire ($ L=0.5~\mu $m) in the presence of strong disorder with $ \sigma_\mu/\mu=3 $. (a)-(d) show the local and nonlocal conductance in the ``intrinsic'' color scale. (e)-(h) are the corresponding line cuts of the conductance as a function of bias at $ V_{\text{Z}}=1.1 $ meV, $ 1.33 $ meV, $ 1.41 $ meV. The other parameters are the same as Fig.~\ref{fig:good}. The corresponding TV from the left (right) and thermal conductance $ \kappa_{RL} $ ($ \kappa_{LR} $) are shown in the inset of (g) [(h)].}
	\label{fig:muVar3S}
\end{figure*}
	
	In Fig.~\ref{fig:muVar3S}, we present the strong disorder with $ \sigma_\mu/\mu=3 $ in the short wire $ L=0.5~\mu $m, as a comparison to its long wire counterpart in Fig.~\ref{fig:muVar3}. The local conductance in Fig.~\ref{fig:muVar3S}(a) shows a trivial ZBCP emerging below the nominal TQPT, and the putative topological ZBCP is destroyed in Fig.~\ref{fig:muVar3S}(b). The nonlocal conductance shows a vague signature of gap closing but without the gap reopening.
	
	In the second row, we present the line cuts of the color plots of the nonlocal conductance from the first row, and find that the nonlocal conductance (blue lines) is also much smaller than the pristine short wire case in Fig.~\ref{fig:good0.5c}. The thermal conductances [red lines in the upper insets of {Figs}.~\ref{fig:muVar3S}(g) and~\ref{fig:muVar3S}(h)] remain finite after the induced gap closes and oscillate as the magnetic field increases. The TV [blue lines in the upper insets of {Figs}.~\ref{fig:muVar3S}(g) and~\ref{fig:muVar3S}(h)] also does not completely flip to -1, which indicates that the topology is not well-defined as expected from the lack of gap reopening seen in Figs.~\ref{fig:muVar3S}{(c) and~\ref{fig:muVar3S}(d)}.

\begin{figure*}[htbp]
	\centering
	\includegraphics[width=6.8in]{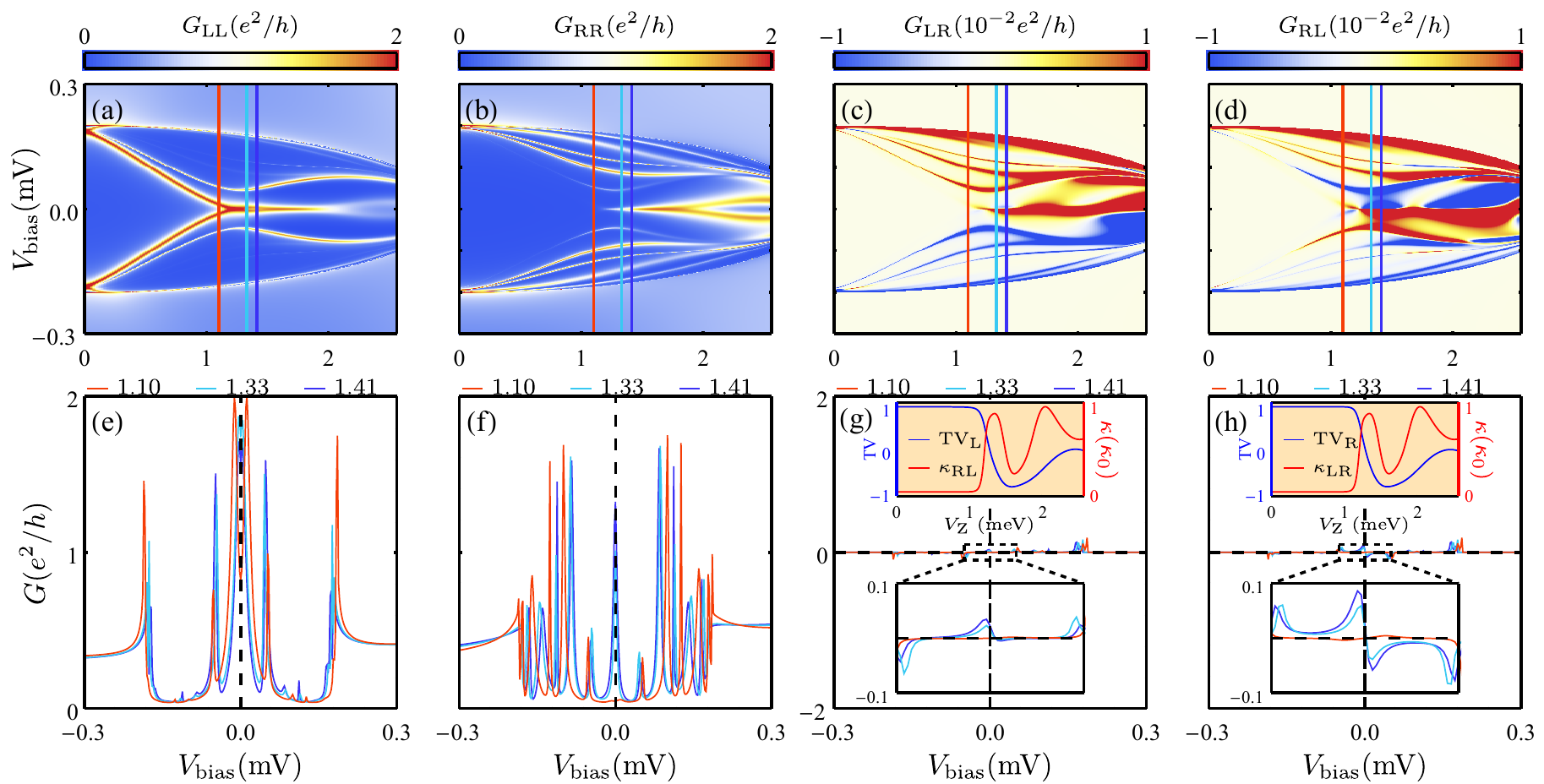}
	\caption{The short wire ($ L=0.5~\mu $m) in the presence of strong disorder with $ \sigma_\mu/\mu=3 $. The SC bulk gap collapses as $ V_{\text{Z}} $ increases and closes completely at $ V_{\text{Z}}=3 $ meV (not shown here). (a)-(d) show the local and nonlocal conductance in the ``intrinsic'' color scale. (e)-(h) are the corresponding line cuts of the conductance as a function of bias at $ V_{\text{Z}}=1.1 $ meV, $ 1.33 $ meV, $ 1.41 $ meV. The other parameters are the same as Fig.~\ref{fig:good}. The corresponding TV from the left (right) and thermal conductance $ \kappa_{RL} $ ($ \kappa_{LR} $) are shown in the inset of (g) [(h)].}
	\label{fig:muVar3Sc} 
\end{figure*}

	Similar to Fig.~\ref{fig:good0.5c}, we show results in a short wire in the presence of strong disorder with the collapse of the parent SC bulk gap in Fig.~\ref{fig:muVar3Sc}. In the local conductance in Figs.~\ref{fig:muVar3Sc}(a) and~\ref{fig:muVar3Sc}(b), we find that the conductances at small magnetic fields are not altered while the conductances at large magnetic fields are more easily affected. In the nonlocal conductance in Figs.~\ref{fig:muVar3Sc}(c) and~\ref{fig:muVar3Sc}(d), the amplitude of the oscillation (anticrossings of two ABS with low energies) above $ V_{\text{Z}}=1.41 $ meV is much suppressed by the collapse of the parent SC gap. 
	
	In the second row, we present the line cuts of the color plots of the nonlocal conductance from the first row, and find that the nonlocal conductance [blue lines in Figs.~\ref{fig:muVar3Sc}(g) and~\ref{fig:muVar3Sc}(h)] near zero bias is almost the same as the one without SC gap collapse [Figs.~\ref{fig:muVar3S}(g) and~\ref{fig:muVar3S}(h)]. The thermal conductances [red lines in the upper insets of Figs.~\ref{fig:muVar3Sc}(g) and~\ref{fig:muVar3Sc}(h)] also remain finite after the induced gap closes, and oscillate as the magnetic field increases. The TV [blue lines in the upper insets of {Figs.}~\ref{fig:muVar3Sc}(g) and~\ref{fig:muVar3Sc}(h)] does not flip to -1 completely, which supports the same conclusion of the ill-defined topology in short wires.

	We add as a sobering reminder that the short-wire and highly disordered situation with the bulk gap collapse appears to be {the most consistent interpretation for the} recent measurements of nonlocal conductance in semiconductor nanowires.~\cite{puglia2020closing} {Obviously, more nonlocal conductance measurements are necessary for progress in the field.}
	\section{Discussion}\label{sec:discussion}
	We discuss now the possible experimental situation and the efficacy of observing MZMs in nanowires based on our results in this paper.
	 
     The local conductance is not a reliable technique to distinguish the ABS from the MBS. Although in the pristine wire in Fig.~\ref{fig:good} the topological ZBCP has a quantized value of $ 2e^2/h $, the opposite assertion is not true. A quantized ZBCP is necessary but not sufficient. In the situation of the quantum dot and inhomogeneous potential in Figs.~\ref{fig:qd} and~\ref{fig:inhom}, the bad ZBCP is also induced on one end of the nanowire by the barrier of the quantum dot or inhomogeneous potential, and it transmutes into the real MBS above TQPT. After the phase transition, the topological regime shows up as the magnetic field increases. This bad ZBCP is sometimes referred to as the quasi-Majorana:\cite{vuik2019reproducing} It generically manifests a quantized and stable conductance peak without oscillation in the trivial regime due to the partial separation of the two Majorana modes,\cite{moore2018twoterminal,stanescu2019robust,moore2018quantized} unlike the ugly ZBCP, where the quantized value of conductance is purely by accident and unstable as a result of the Anderson localization.~\cite{anderson1958absence} Thus, the local conductance is not sufficient to confirm the MBS. This has already been discussed in the literature.
     
     The nonlocal conductance should be more informative because it contains the bulk information. In the good ZBCP, the conductances at the bias above the parent gap and below the proximity gap are nearly zero; therefore, we can directly observe the {gap closure} and gap reopening in the nonlocal conductance. However, in the bad ZBCP in Figs.~\ref{fig:qd} and~\ref{fig:inhom}, the nonlocal conductance is too small to observe any signature near zero bias, which restates the need for {a} very high precision in the three-terminal measurement. This situation becomes worse in the ugly ZBCP in the presence of disorder. Disorder generally suppresses the magnitude of the nonlocal conductance everywhere. For very strong disorder, the topological regime is not well defined; the system just enters the Anderson localization regime. Therefore, the discussion of topology in such a case is totally meaningless since the wire is just composed of a bunch of quantum dots that cannot be described by the nanowire model effectively. Thus, the nonlocal conductance will not affirmatively tell us whether a system is in the topological regime or not; on the other hand, if the very weak signal is detected in experiments, it may indicate that the underlying disorder is very strong and the ZBCP in the local conductance, if any, would be the ugly ZBCP induced by disorder. The invariable presence of noise in experiments may become a real challenge in the context of detecting such weak nonlocal signals.
     
     The thermal conductance provides more reliability in determining the topological regime. In the good ZBCP, the thermal conductance sharply peaks at the vanishing point of TV, which retains a quantized thermal conductance at that point. This peak does not go away even if we add the inhomogeneous potential and quantum dot, although the quantization may not manifest due to finite dissipation. In the presence of disorder, the thermal conductance is still quite robust, though not quantized due to the combined effect of finite dissipation and disorder. Of course, the thermal conductance cannot be helpful in the very strong disorder limit since the topology itself is not well defined--- we {do not} see a peak of thermal conductance in this situation. Other than this strong disorder {situation}, we would consider the thermal conductance as a robust and accurate indicator of the TQPT. On the other hand, if we always find the negligible thermal conductance when sweeping the magnetic field, it indicates the disorder to be so large that the topological regime may not even exist. Nothing useful will happen in experiments if the samples are in the strong disorder regime since all one is exploring is the physics of Anderson localization, and not the physics of topological SC.
	 
	 From the good ZBCP in the pristine wire to the bad ZBCP in the inhomogeneous potential and quantum dot, and to ugly ZBCP in the presence of disorder, we conclude that the local conductance is useless in distinguishing the ABS from the MBS; the nonlocal conductance--- which may provide more bulk information--- is in principle useful, but is unfortunately more fragile to the inhomogeneous potential and disorder; therefore, it may not serve as a very practical tool. However, the thermal conductance is a more reliable technique to predict the topology of the nanowire.
	
	 In short wires, all the conductances--- local, nonlocal or thermal--- are not illuminating since the topological regime itself is not well-defined from a practical viewpoint. Therefore, we emphasize that all the discussion of the electrical and thermal conductance above should be considered only in the context of the long wire limit.
	 
	\section{Conclusion}\label{sec:conclusion}
	
	\begin{table}[ht]
		
		\begin{ruledtabular}
			\caption{Summary for different measurements in different situations.}
			\begin{tabular}{lcccr}
				 & $ G_{ii} $ &  $ G_{ij} $ & $ \kappa $ & TV\\
				\hline
				Good-pristine & $\checkmark$ & $\checkmark$ & $\checkmark$ & $\checkmark$ \\
				Bad-QD & $ \times $ & $\checkmark$ & $\checkmark$ & $\checkmark$ \\
				Bad-inhomogeneous potential& $ \times $ & $\checkmark^*$ &$\checkmark$&$\checkmark$\\
				Ugly-intermediate disorder & $ \times $ & $\checkmark^*$ & $\checkmark$ & $\checkmark$\\
				Ugly-strong disorder &$ \times $  & $ \times $ &$ \times $ & $ \times $\\	
				Short wires &$ \times $  & $ \times $ &$ \times $ & $ \times $
			\end{tabular}
		
		\label{tab:summary}
		\end{ruledtabular}
	\end{table}

	In this paper, we exhaustively simulate the electrical and thermal conductance in the three-terminal semiconductor-superconductor hybrid Majorana nanowire device in various situations, including pristine nanowire, quantum dots, inhomogeneous potential, and onsite disorder in the chemical potential. The goal is to provide extensive results so that forthcoming experimental nonlocal conductance data can be validated through a direct comparison with our theory.   To better simulate experiments, we also change the long wire limit to the short wire and introduce the parent SC bulk gap collapse at a high magnetic field. We summarize the efficacy of all four measurements (although the topological visibility is only theoretically measured) to distinguish each situation in Table~\ref{tab:summary}.  The check $\checkmark$ (cross $ \times $) means a possible (impossible) measurement to distinguish a particular situation. The extra star $ * $ after the check $ \checkmark $ indicates a possible measurement but with a significantly weaker signal compared to the pristine case. The four columns are local conductance $ G_{ii} $, nonlocal conductance $ G_{ij} $, thermal conductance $ \kappa $ , and the topological visibility (TV).
	
	This work generalizes our previous paper~\cite{pan2020physical} on the good, bad, and ugly ZBCP in the local conductance to the nonlocal conductance and thermal conductance. We first present the nonlocal conductance in the pristine nanowire, which results in the good ZBCP, to give a general idea of what the nonlocal conductance will look like if it is the real topological Majorana in a clean long wire. This serves as a guide for future experiments which may report the observation of topological Majorana in nonlocal conductance. The local conductance has a quantized peak at $ 2e^2/h $ in the topological regime; the nonlocal conductance shows an apparent {gap closure} and reopening signal at TQPT, but in general the signal is much weaker than the local conductance. This TQPT, which is determined by the sign flip of topological visibility, can be captured clearly by the peak of thermal conductance. It may therefore be necessary to carry out local conductance, nonlocal conductance, and thermal conductance measurements in the same sample for conclusive results.  This becomes particularly significant in view of disorder considerably complicating and weakening the nonlocal conductance signal.
	
	We also investigate the bad ZBCP arising from inhomogeneous potential including quantum dots. This may induce the quasi-Majorana Andreev bound state in the trivial regime with a quantized ZBCP. As the magnetic field increases, the trivial bad ZBCP finally transmutes into the topological ZBCP. We find that the local conductance here becomes misleading and the nonlocal conductance, while still showing the {gap closure} and reopening feature near TQPT, has much weaker signals than the local conductance. It is unclear if such a weak signal is experimentally detectable in the presence of noise, but such measurements are {necessary} for progress since these nonlocal measurements are capable of detecting bulk topological properties. However, the peak of the thermal conductance can still provide evidence for the real TQPT. Thus, measuring thermal conductance along with nonlocal conductance is warranted for decisive conclusions.
	
	Most relevant to experiments, we consider the role of disorder in nonlocal conductance.  This is crucial since, based on our earlier work,~\cite{pan2020physical} we believe that the existing NS tunneling, measurements may possibly be observing disorder induced ZBCP effects. We study a nanowire in the presence of disorder in the chemical potential, and find that weak disorder does not destroy topological effects: The electrical conductance and thermal conductance in the weak disorder resembles that in the pristine wire. In the presence of the intermediate disorder, we find that the trivial ZBCP emerges in the local conductance below the putative TQPT which is misleading if one decides only based on the local conductance (since local conductance measurement knows nothing about TQPT and there is no way to tell whether a ZBCP is above or below the TQPT based only on local conductance data). A decisive measurement of three-terminal conductance as proposed in this work may enable the observation of bulk topological properties, but one must be careful about the weak strength of the nonlocal signal. However, if we measure the thermal conductance, we notice that the thermal conductance still peaks at the real TQPT, which can be determined by the zero of topological visibility. The only difference is that the range of the topological regime now becomes small compared to the pristine case. The worst case is strong disorder, which completely destroys the topological regime. The system is dominated by the strong disorder (and not superconductivity) and enters the Anderson localization regime. Here the local conductance is purely random as in the class D ensemble, the nonlocal is very small, and thermal conductance shows irregular {structures} with many peaks. Thus, it is {not meaningful} to discuss the topological or nontopological in such a strong disorder situation, and the only possible conclusion then would be that the sample quality must improve before MZMs can manifest.
	
	We also consider a realistic situation by combining disorder and quantum dot. We find two limits based on the magnitude of disorder. For the weak disorder, the nanowire is still dominated by the quantum dot, which manifests the bad ZBCP as if the disorder is absent. For the strong disorder, the nanowire is dominated by disorder so that the bad ZBCP becomes the ugly ZBCP. The local conductance now becomes arbitrary, and the putative topological regime disappears.
	
	Finally, simulating the current experimental short wire situations, we perform the same calculation in a short wire ($ L=0.5~\mu $m) and introduce the phenomenological parent SC bulk gap collapse which happens in the experiment. In the short wire, there is no distinction between the topological and trivial regime from a practical standpoint since the topology itself is not well defined when the wire is much shorter than the coherence length. The overlap of Majorana from both ends in the short wire is so large that the thermal conductance remains finite even after the proximity gap closes. This is a situation that should be avoided in experiments, although we believe that all experiments so far, unfortunately, may be in this undesirable short wire limit.
	
	Our extensive results presented in this work establish that nonlocal three-terminal conductance along with the standard two-terminal tunneling spectroscopy could distinguish between topological and trivial regimes in Majorana nanowires provided the wires are not too short and not too dirty, in particular, any observation of {gap closure} and reopening features in the nonlocal conductance, even if the signal is weak, along with a stable quantized tunnel conductance peak should be {reasonable} evidence for topological Majorana zero modes.  If thermal conductance measurements can also be carried out simultaneously and manifest our predicted signatures, one could be {more assured} that the nanowire is indeed in the topological regime, and one could then proceed toward the goal of braiding the Majorana modes in order to create a protected topological qubit.
	
	This work is supported by Laboratory for Physical Sciences and Microsoft. We also acknowledge the support of the University of Maryland supercomputing resources (hpcc.umd.edu).

	\bibliography{3terminal}
	\onecolumngrid
	\appendix
	\setcounter{secnumdepth}{3}
	\renewcommand\figurename{Figure}

	\section{Other examples of the good, bad, and ugly ZBCPs}\label{app:A}
\begin{figure}[h]
	\centering
	\includegraphics[width=6.8in]{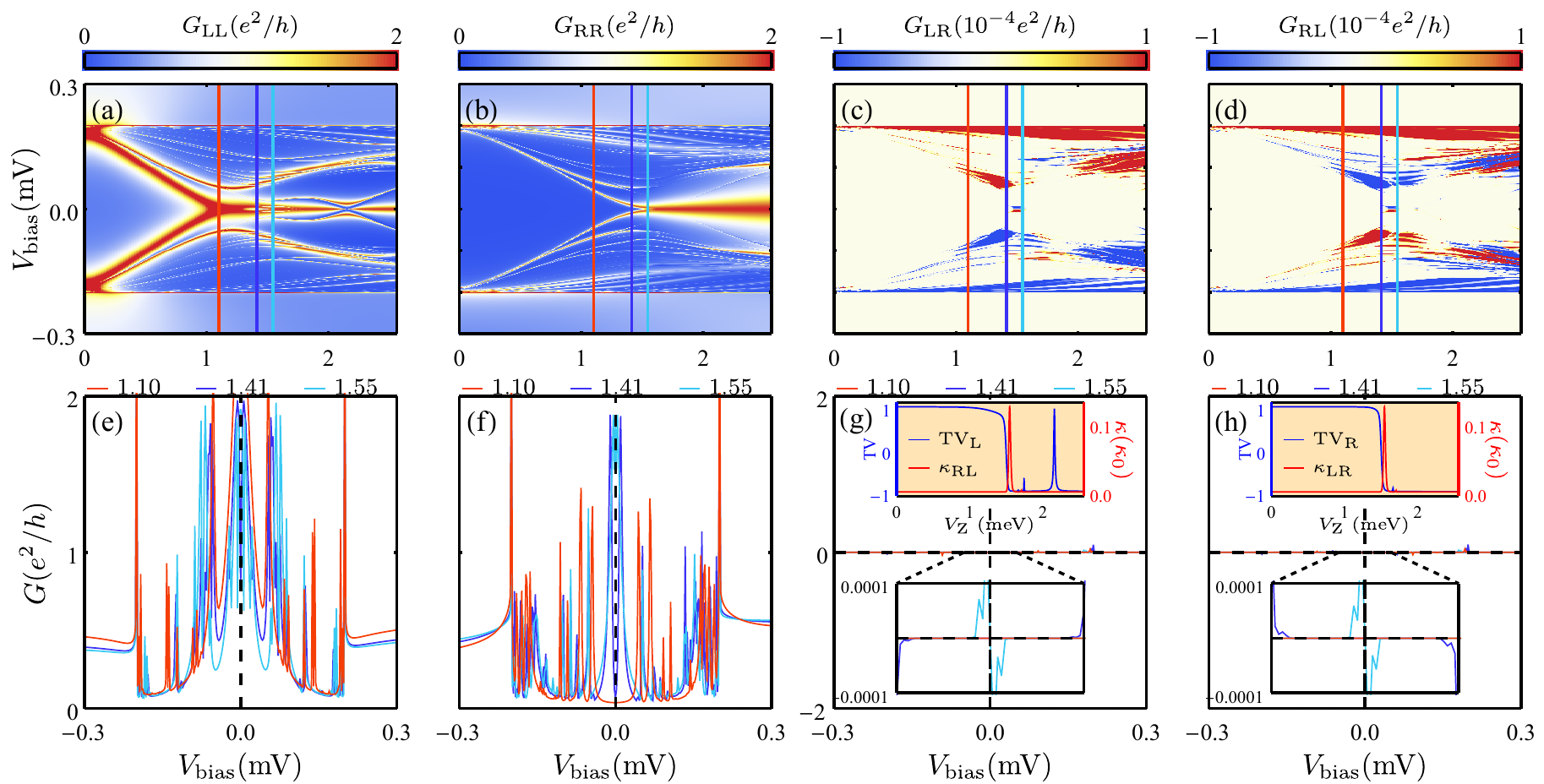}
	\caption{The ugly ZBCP in a nanowire in the presence of intermediate disorder with $ \sigma_\mu/\mu=2 $. (a)-(d) show the local and nonlocal conductance in the ``intrinsic'' color scale. (e)-(h) are the corresponding line cuts of the conductance as a function of bias at $ V_{\text{Z}}=1.1 $ meV, $ 1.41 $ meV, $ 1.55 $ meV. The other parameters are the same as Fig.~\ref{fig:good}. The corresponding TV from the left (right) and thermal conductance $ \kappa_{RL} $ ($ \kappa_{LR} $) are shown in the inset of (g) [(h)].}
	\label{fig:muVar2}
	\end{figure}

\begin{figure}[h]
	\centering
	\includegraphics[width=6.8in]{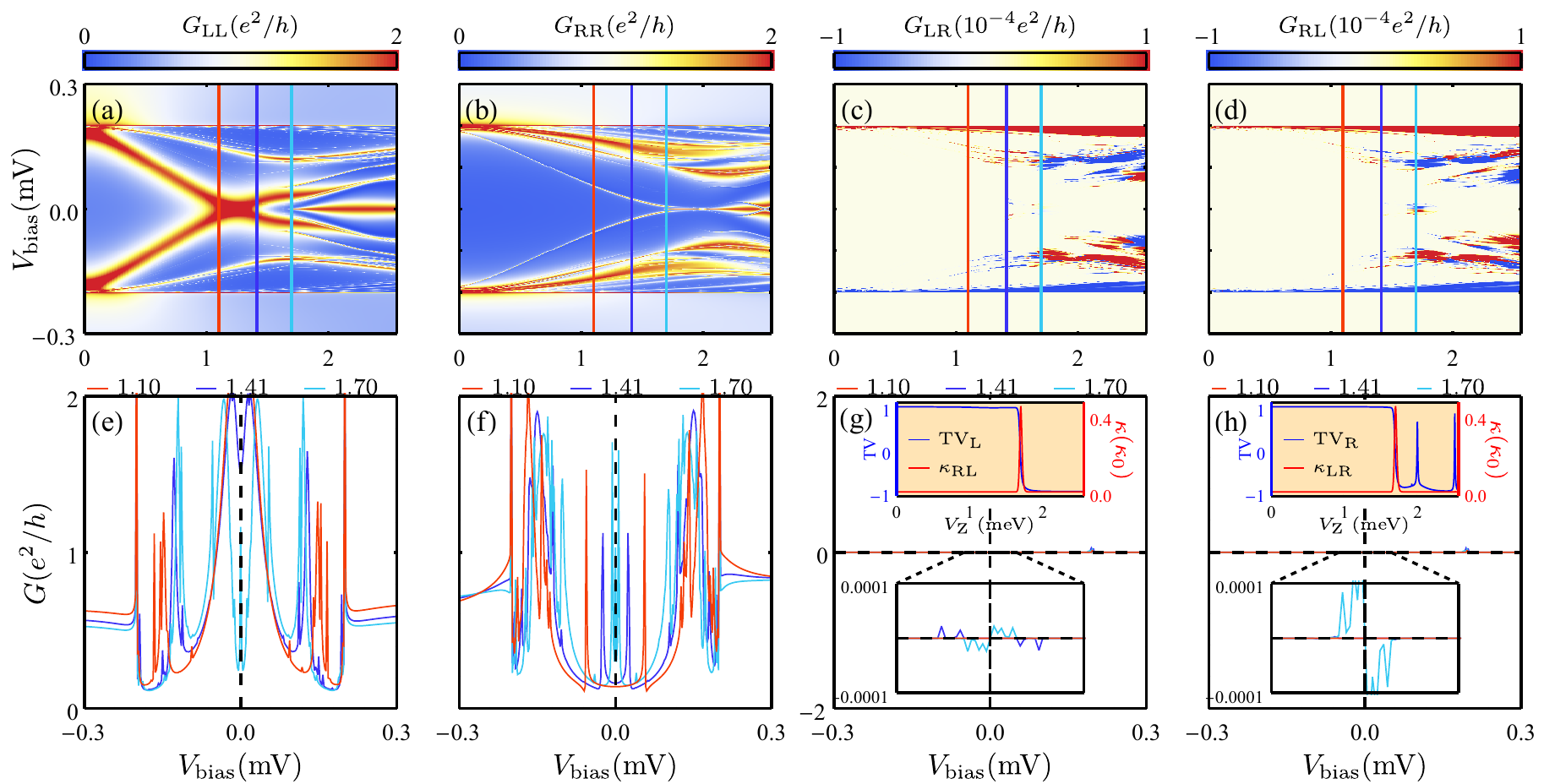}
	\caption{The ugly ZBCP in a nanowire in the presence of intermediate disorder with $ \sigma_\mu/\mu=2.5 $. (a)-(d) show the local and nonlocal conductance in the ``intrinsic'' color scale. (e)-(h) are the corresponding line cuts of the conductance as a function of bias at $ V_{\text{Z}}=1.1 $ meV, $ 1.41 $ meV, $ 1.7 $ meV. The other parameters are the same as Fig.~\ref{fig:good}. The corresponding TV from the left (right) and thermal conductance $ \kappa_{RL} $ ($ \kappa_{LR} $) are shown in the inset of (g) [(h)].}
	\label{fig:muVar2p5}
	\end{figure}

\begin{figure}[h]
	\centering
	\includegraphics[width=6.8in]{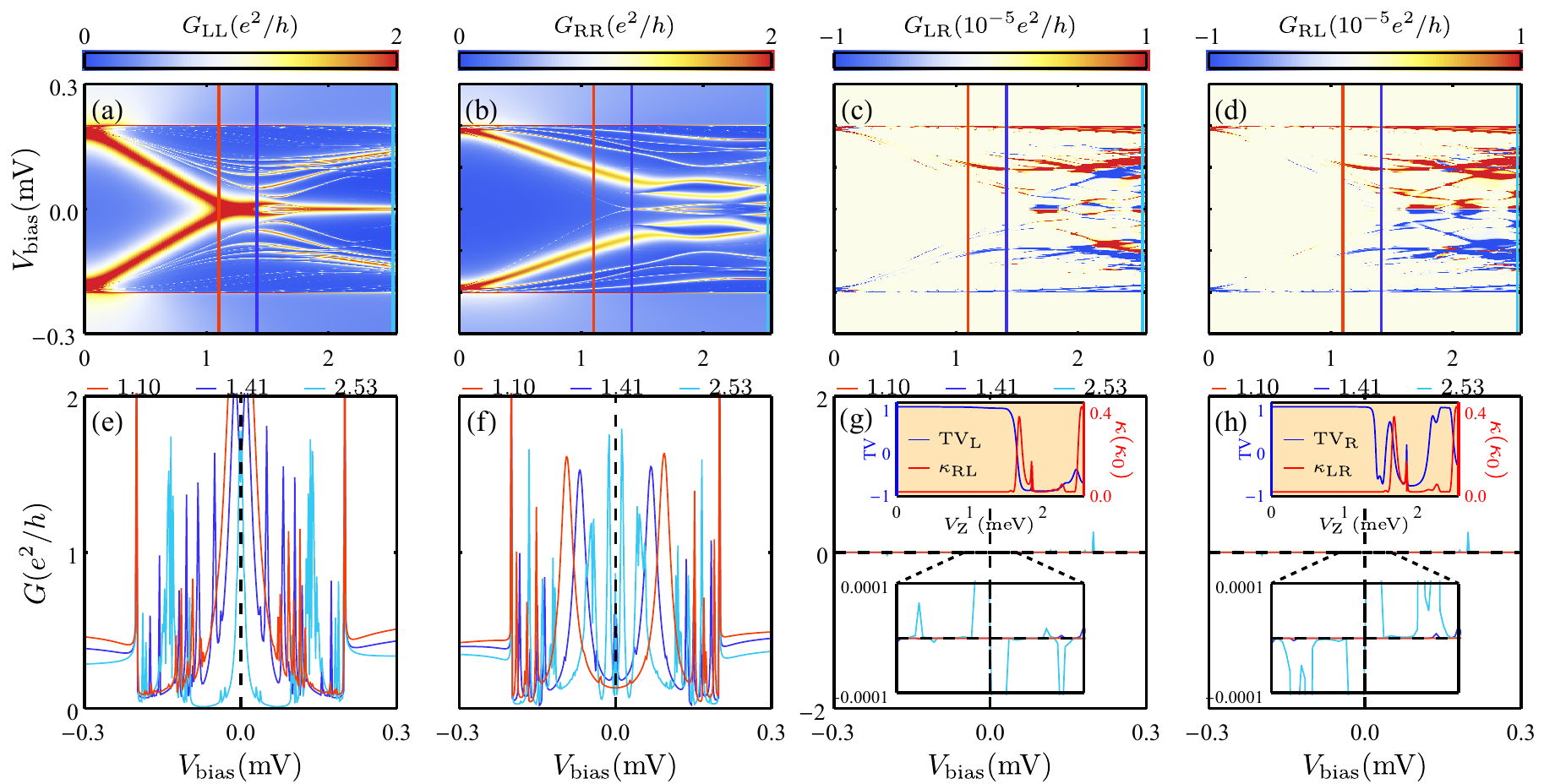}
	\caption{The ugly ZBCP in a nanowire in the presence of strong disorder with $ \sigma_\mu/\mu=3.5 $. (a)-(d) show the local and nonlocal conductance in the ``intrinsic'' color scale. (e)-(h) are the corresponding line cuts of the conductance as a function of bias at $ V_{\text{Z}}=1.1 $ meV, $ 1.41 $ meV, $ 2.53$ meV. The other parameters are the same as Fig.~\ref{fig:good}. The corresponding TV from the left (right) and thermal conductance $ \kappa_{RL} $ ($ \kappa_{LR} $) are shown in the inset of (g) [(h)].}
	\label{fig:muVar3p5}
	\end{figure}

\begin{figure}[h]
		\centering
		\includegraphics[width=6.8in]{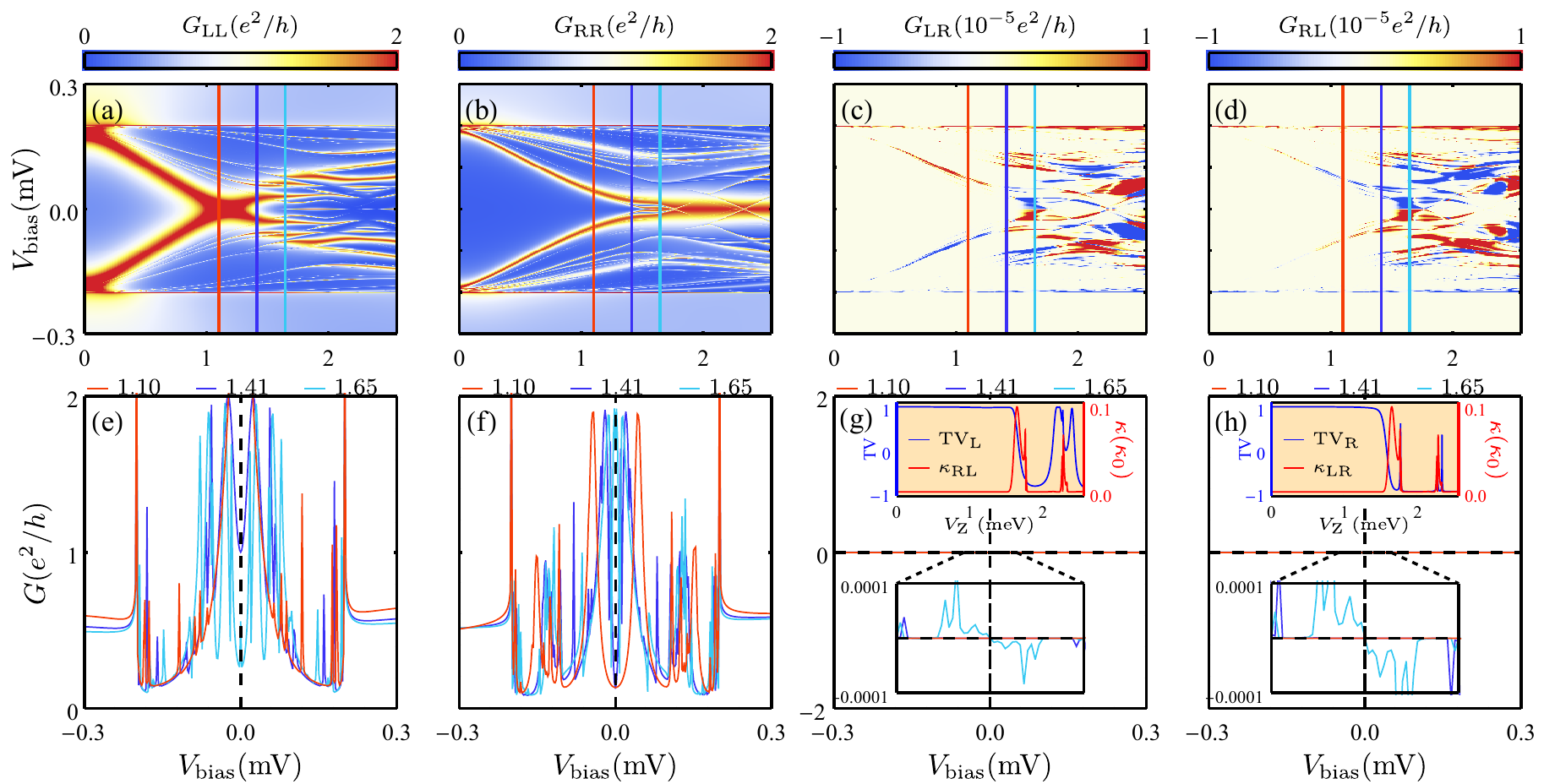}
		\caption{The ugly ZBCP in a nanowire in the presence of strong disorder with $ \sigma_\mu/\mu=4 $. (a)-(d) show the local and nonlocal conductance in the ``intrinsic'' color scale. (e)-(h) are the corresponding line cuts of the conductance as a function of bias at $ V_{\text{Z}}=1.1 $ meV, $ 1.41 $ meV, $ 1.65 $ meV. The other parameters are the same as Fig.~\ref{fig:good}. The corresponding TV from the left (right) and thermal conductance $ \kappa_{RL} $ ($ \kappa_{LR} $) are shown in the inset of (g) [(h)].}
		\label{fig:muVar4}
	\end{figure}

\begin{figure}[h]
		\centering
		\includegraphics[width=6.8in]{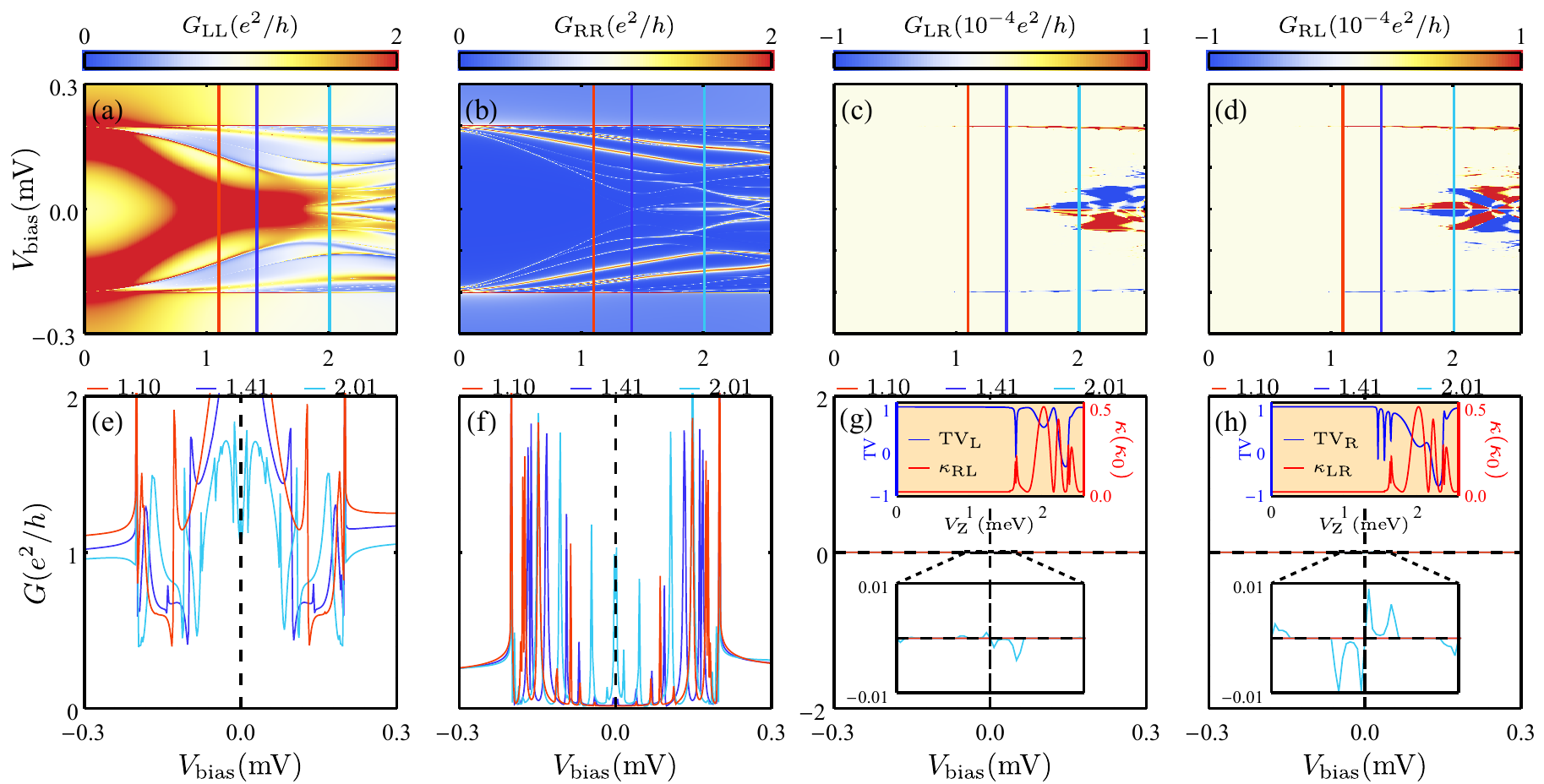}
		\caption{The ugly ZBCP in a nanowire in the presence of strong disorder with $ \sigma_\mu/\mu=4.5 $. (a)-(d) show the local and nonlocal conductance in the ``intrinsic'' color scale. (e)-(h) are the corresponding line cuts of the conductance as a function of bias at $ V_{\text{Z}}=1.1 $ meV, $ 1.41 $ meV, $ 2.01 $ meV. The other parameters are the same as Fig.~\ref{fig:good}. The corresponding TV from the left (right) and thermal conductance $ \kappa_{RL} $ ($ \kappa_{LR} $) are shown in the inset of (g) [(h)].}
		\label{fig:muVar4p5}
	\end{figure}

\begin{figure}[h]
		\centering
		\includegraphics[width=6.8in]{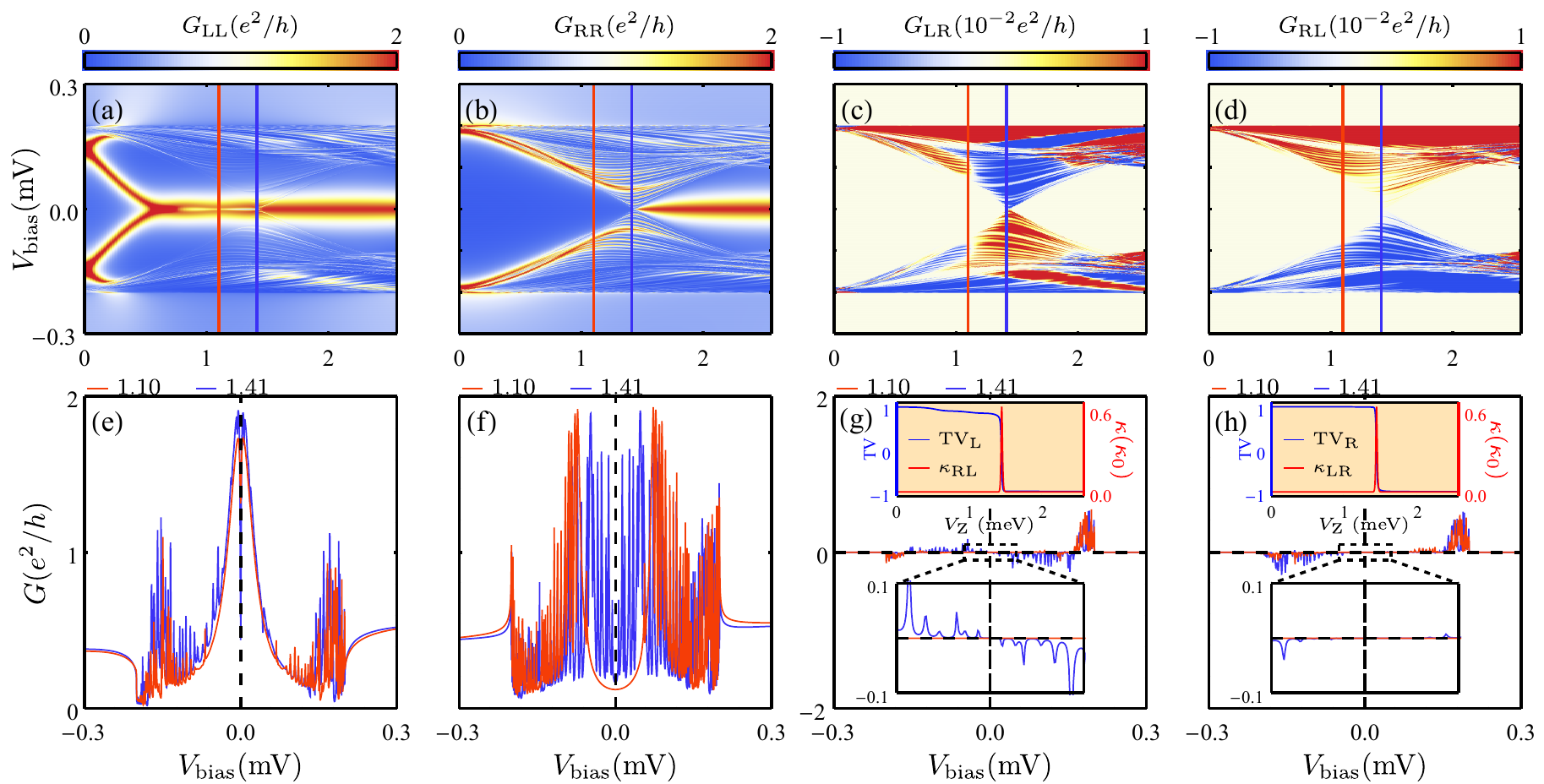}
		\caption{The wire in the presence of the quantum dot and weak disorder with $ \sigma_\mu/\mu=0.5 $. (a)-(d) show the local and nonlocal conductance in the ``intrinsic'' color scale. (e)-(h) are the corresponding line cuts of the conductance as a function of bias at $ V_{\text{Z}}=1.1 $ meV, and $ 1.41 $ meV. The quantum dot is the same as Fig.~\ref{fig:qd} with $ V_{\text{D}}=0.4 $ meV and $ l=0.15~\mu $m. The other parameters are the same as Fig.~\ref{fig:good}. The corresponding TV from the left (right) and thermal conductance $ \kappa_{RL} $ ($ \kappa_{LR} $) are shown in the inset of (g) [(h)].}
		\label{fig:qdmuVar0p5}
	\end{figure}

\begin{figure}[h]
		\centering
		\includegraphics[width=6.8in]{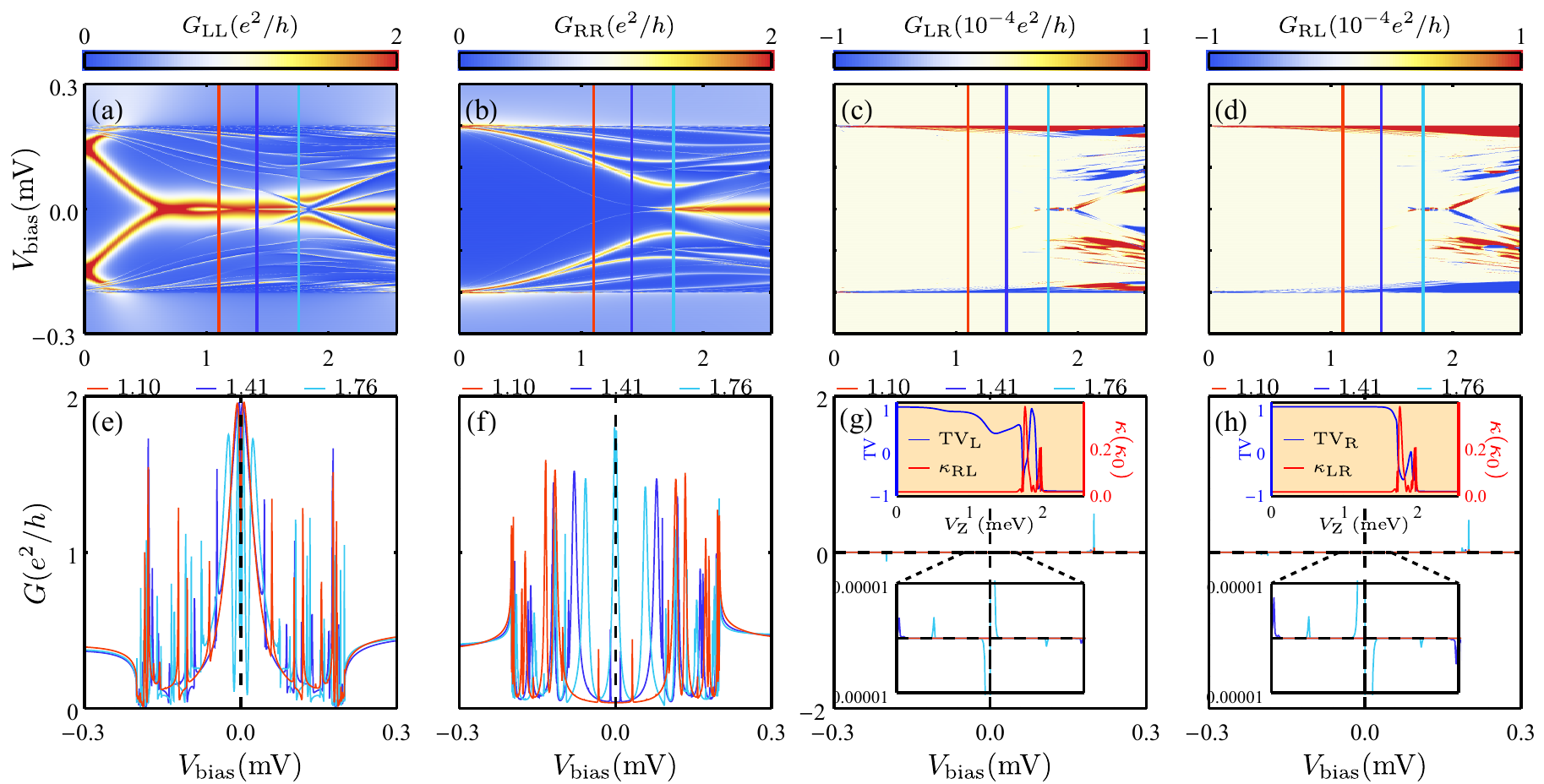}
		\caption{The wire in the presence of the quantum dot and intermediate disorder with $ \sigma_\mu/\mu=3 $. (a)-(d) show the local and nonlocal conductance in the ``intrinsic'' color scale. (e)-(h) are the corresponding line cuts of the conductance as a function of bias at $ V_{\text{Z}}=1.1 $ meV, $ 1.41 $, and $ 1.76 $ meV. The quantum dot is the same as Fig.~\ref{fig:qd} with $ V_{\text{D}}=0.4 $ meV and $ l=0.15~\mu $m. The other parameters are the same as Fig.~\ref{fig:good}. The corresponding TV from the left (right) and thermal conductance $ \kappa_{RL} $ ($ \kappa_{LR} $) are shown in the inset of (g) [(h)].}
		\label{fig:qdmuVar3}
	\end{figure}

\begin{figure}[h]
		\centering
		\includegraphics[width=6.8in]{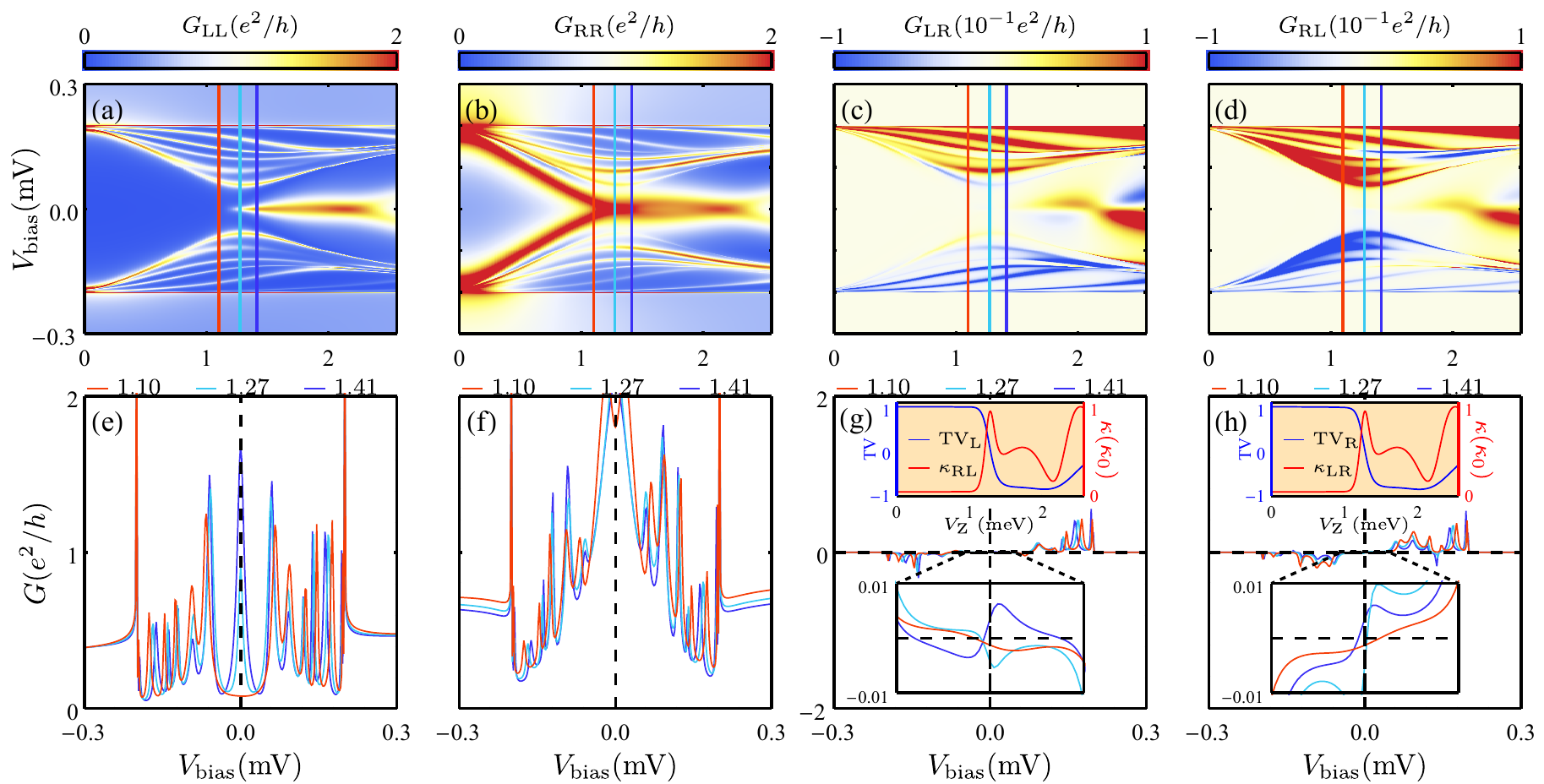}
		\caption{The short wire ($ L=0.5~\mu $m) in the presence of intermediate disorder with $ \sigma_\mu/\mu=1.5 $. (a)-(d) show the local and nonlocal conductance in the ``intrinsic'' color scale. (e)-(h) are the corresponding line cuts of the conductance as a function of bias at $ V_{\text{Z}}=1.1 $ meV, $ 1.27 $ meV, $ 1.41 $ meV. The other parameters are the same as Fig.~\ref{fig:good}. The corresponding TV from the left (right) and thermal conductance $ \kappa_{RL} $ ($ \kappa_{LR} $) are shown in the inset of (g) [(h)].}
		\label{fig:muVar1p5S}
	\end{figure}

\begin{figure}[h]
		\centering
		\includegraphics[width=6.8in]{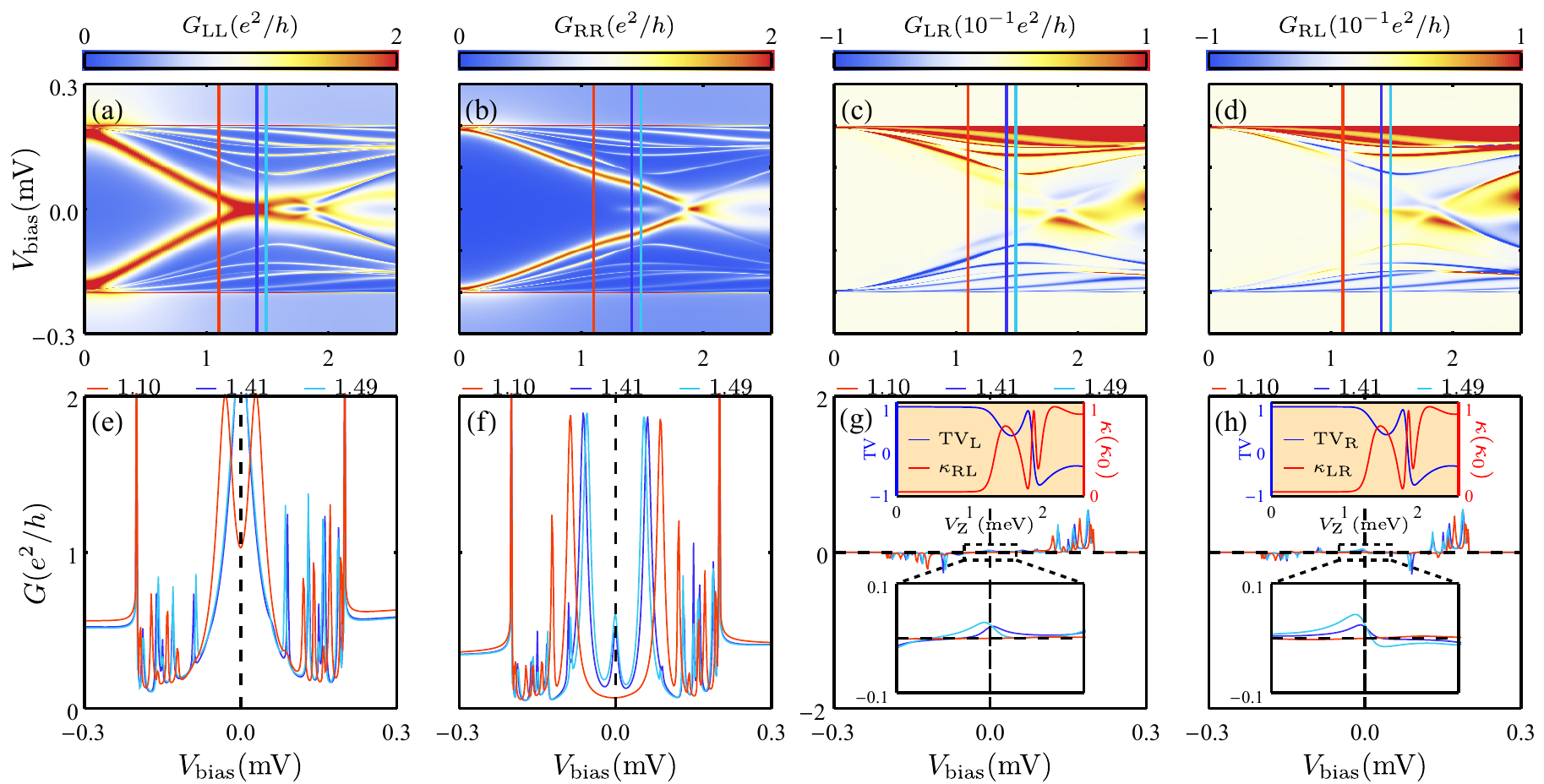}
		\caption{The short wire ($ L=0.5~\mu $m) in the presence of intermediate disorder with $ \sigma_\mu/\mu=2 $. (a)-(d) show the local and nonlocal conductance in the ``intrinsic'' color scale. (e)-(h) are the corresponding line cuts of the conductance as a function of bias at $ V_{\text{Z}}=1.1 $ meV, $ 1.41 $ meV, $ 1.49 $ meV. The other parameters are the same as Fig.~\ref{fig:good}. The corresponding TV from the left (right) and thermal conductance $ \kappa_{RL} $ ($ \kappa_{LR} $) are shown in the inset of (g) [(h)].}
		\label{fig:muVar2S}
	\end{figure}

\begin{figure}[h]
	\centering
	\includegraphics[width=6.8in]{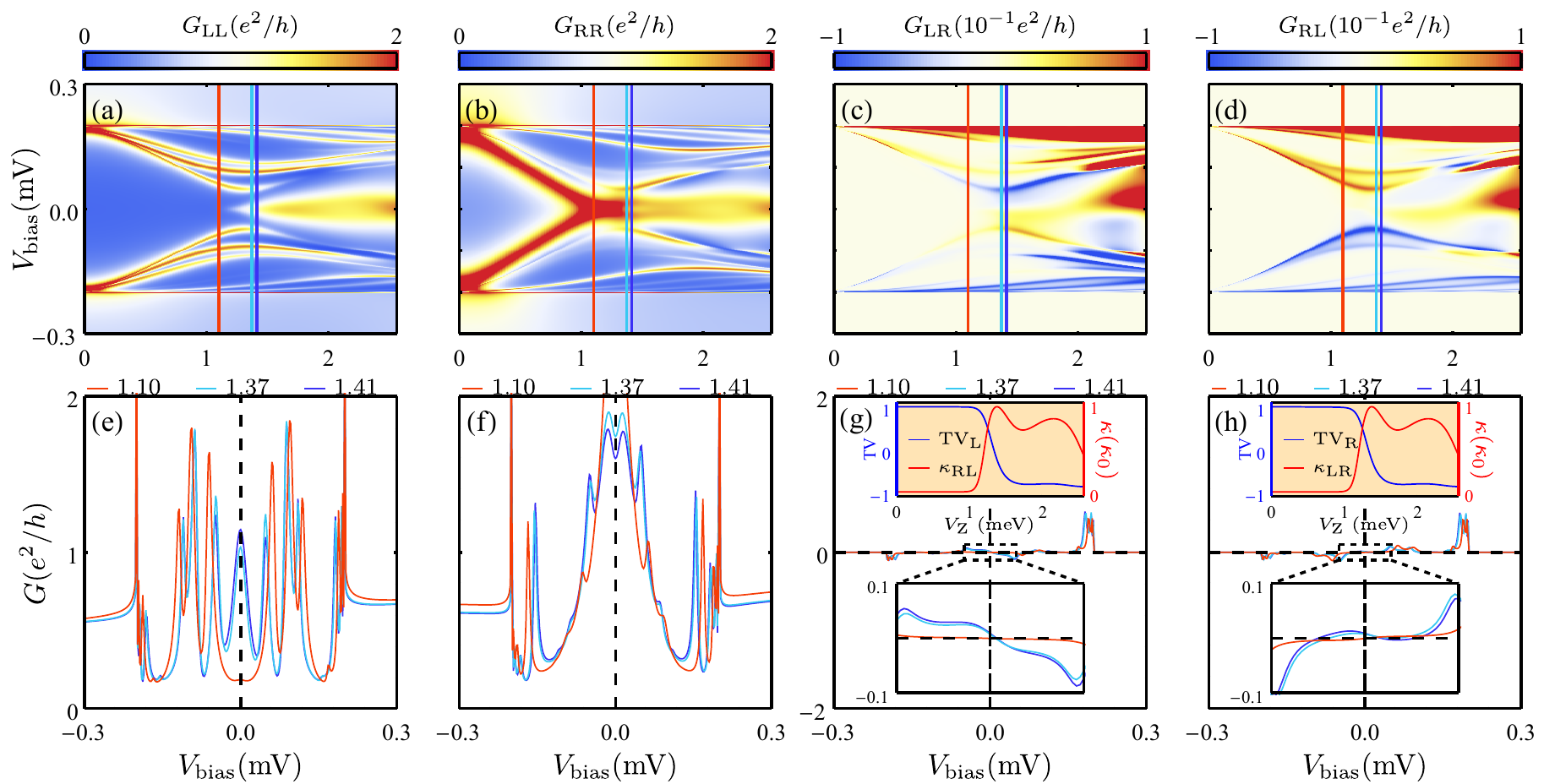}
	\caption{The short wire ($ L=0.5~\mu $m) in the presence of intermediate disorder with $ \sigma_\mu/\mu=2.5 $. (a)-(d) show the local and nonlocal conductance in the ``intrinsic'' color scale. (e)-(h) are the corresponding line cuts of the conductance as a function of bias at $ V_{\text{Z}}=1.1 $ meV, $ 1.37 $ meV, $ 1.41 $ meV. The other parameters are the same as Fig.~\ref{fig:good}. The corresponding TV from the left (right) and thermal conductance $ \kappa_{RL} $ ($ \kappa_{LR} $) are shown in the inset of (g) [(h)].}
	\label{fig:muVar2p5S}
\end{figure}

\begin{figure}[h]
	\centering
	\includegraphics[width=6.8in]{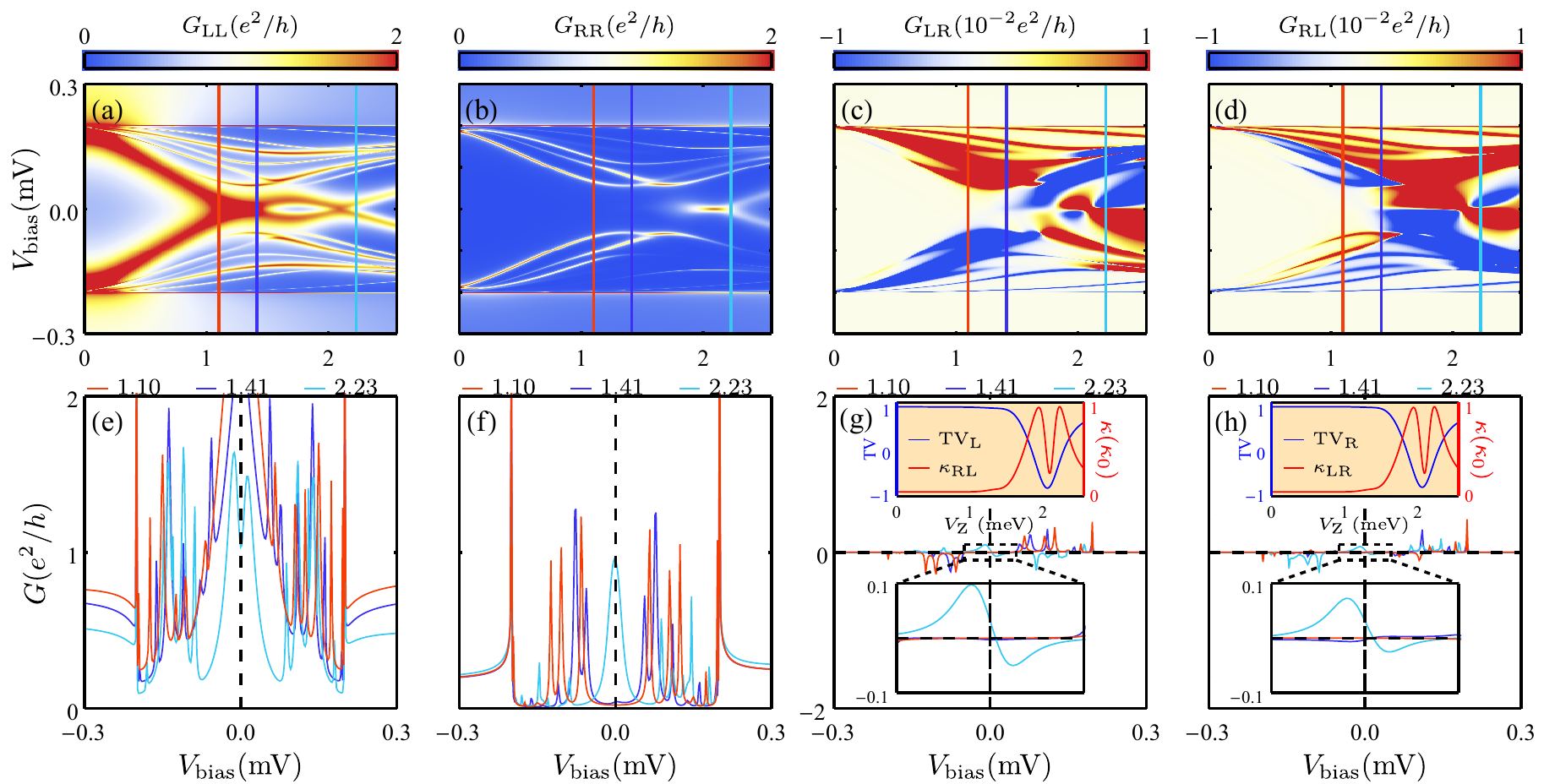}
	\caption{The short wire ($ L=0.5~\mu $m) in the presence of strong disorder with $ \sigma_\mu/\mu=3.5 $. (a)-(d) show the local and nonlocal conductance in the ``intrinsic'' color scale. (e)-(h) are the corresponding line cuts of the conductance as a function of bias at $ V_{\text{Z}}=1.1 $ meV, $ 1.41 $ meV, $ 2.23 $ meV. The other parameters are the same as Fig.~\ref{fig:good}. The corresponding TV from the left (right) and thermal conductance $ \kappa_{RL} $ ($ \kappa_{LR} $) are shown in the inset of (g) [(h)].}
	\label{fig:muVar3p5S}
\end{figure}

\begin{figure}[h]
	\centering
	\includegraphics[width=6.8in]{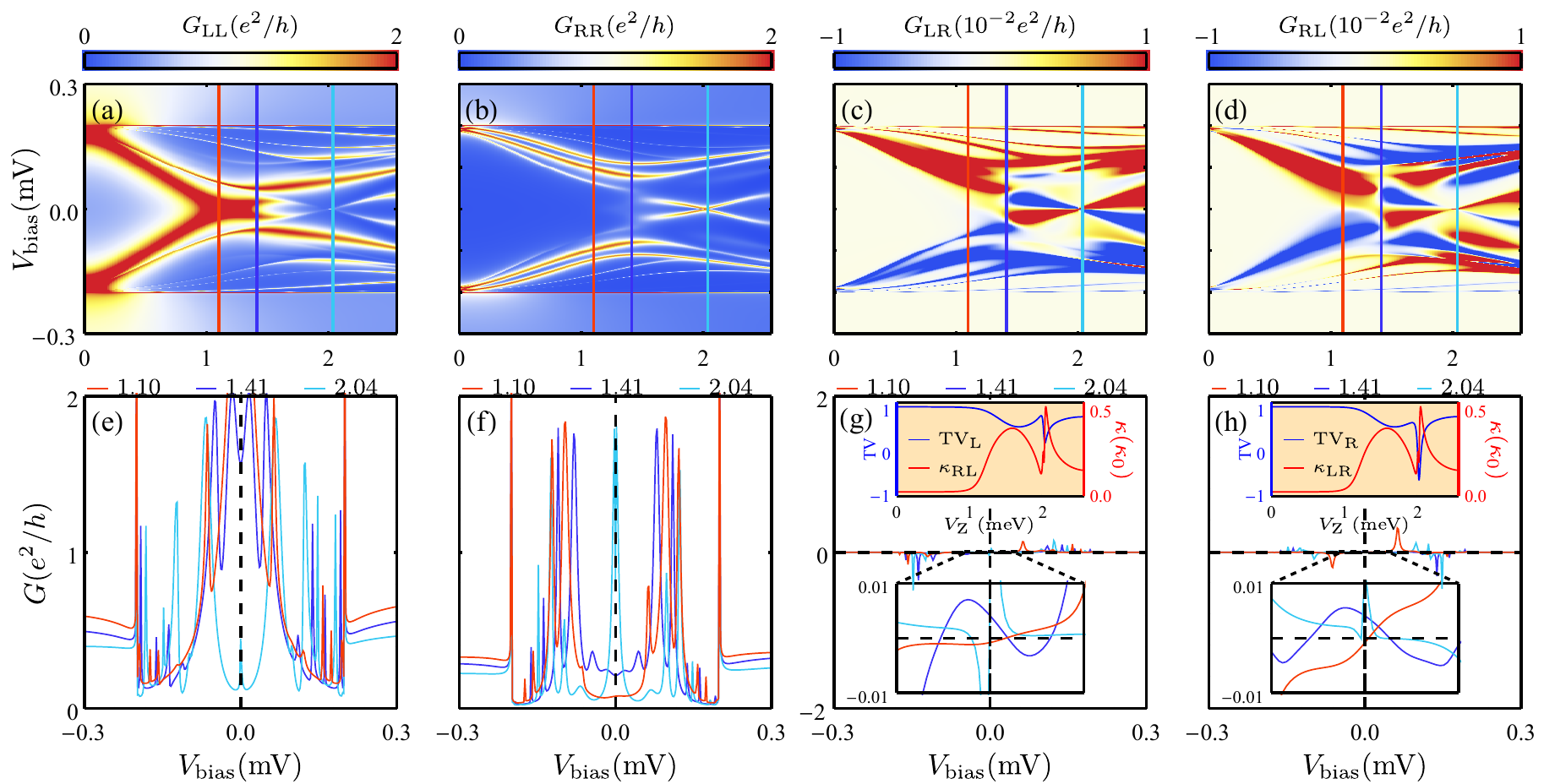}
	\caption{The short wire ($ L=0.5~\mu $m) in the presence of strong disorder with $ \sigma_\mu/\mu=4 $. (a)-(d) show the local and nonlocal conductance in the ``intrinsic'' color scale. (e)-(h) are the corresponding line cuts of the conductance as a function of bias at $ V_{\text{Z}}=1.1 $ meV, $ 1.41 $ meV, $ 2.04 $ meV. The other parameters are the same as Fig.~\ref{fig:good}. The corresponding TV from the left (right) and thermal conductance $ \kappa_{RL} $ ($ \kappa_{LR} $) are shown in the inset of (g) [(h)].}
	\label{fig:muVar4S}
\end{figure}

\begin{figure}[h]
	\centering
	\includegraphics[width=6.8in]{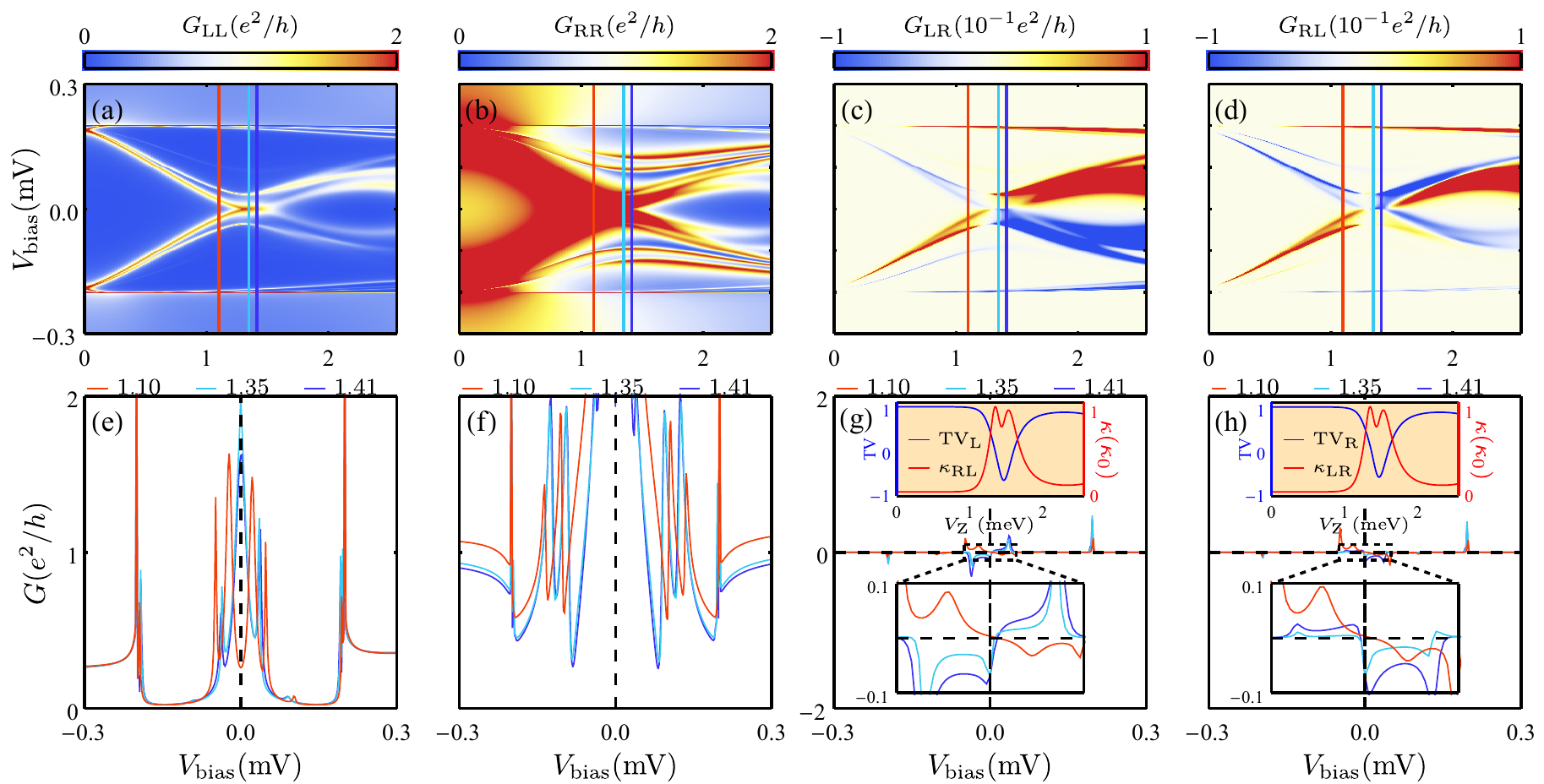}
	\caption{The short wire ($ L=0.5~\mu $m) in the presence of strong disorder with $ \sigma_\mu/\mu=4.5 $. (a)-(d) show the local and nonlocal conductance in the ``intrinsic'' color scale. (e)-(h) are the corresponding line cuts of the conductance as a function of bias at $ V_{\text{Z}}=1.1 $ meV, $ 1.35 $ meV, $ 1.41 $ meV. The other parameters are the same as Fig.~\ref{fig:good}. The corresponding TV from the left (right) and thermal conductance $ \kappa_{RL} $ ($ \kappa_{LR} $) are shown in the inset of (g) [(h)].}
	\label{fig:muVar4p5S}
\end{figure}

\begin{figure}[h]
	\centering
	\includegraphics[width=6.8in]{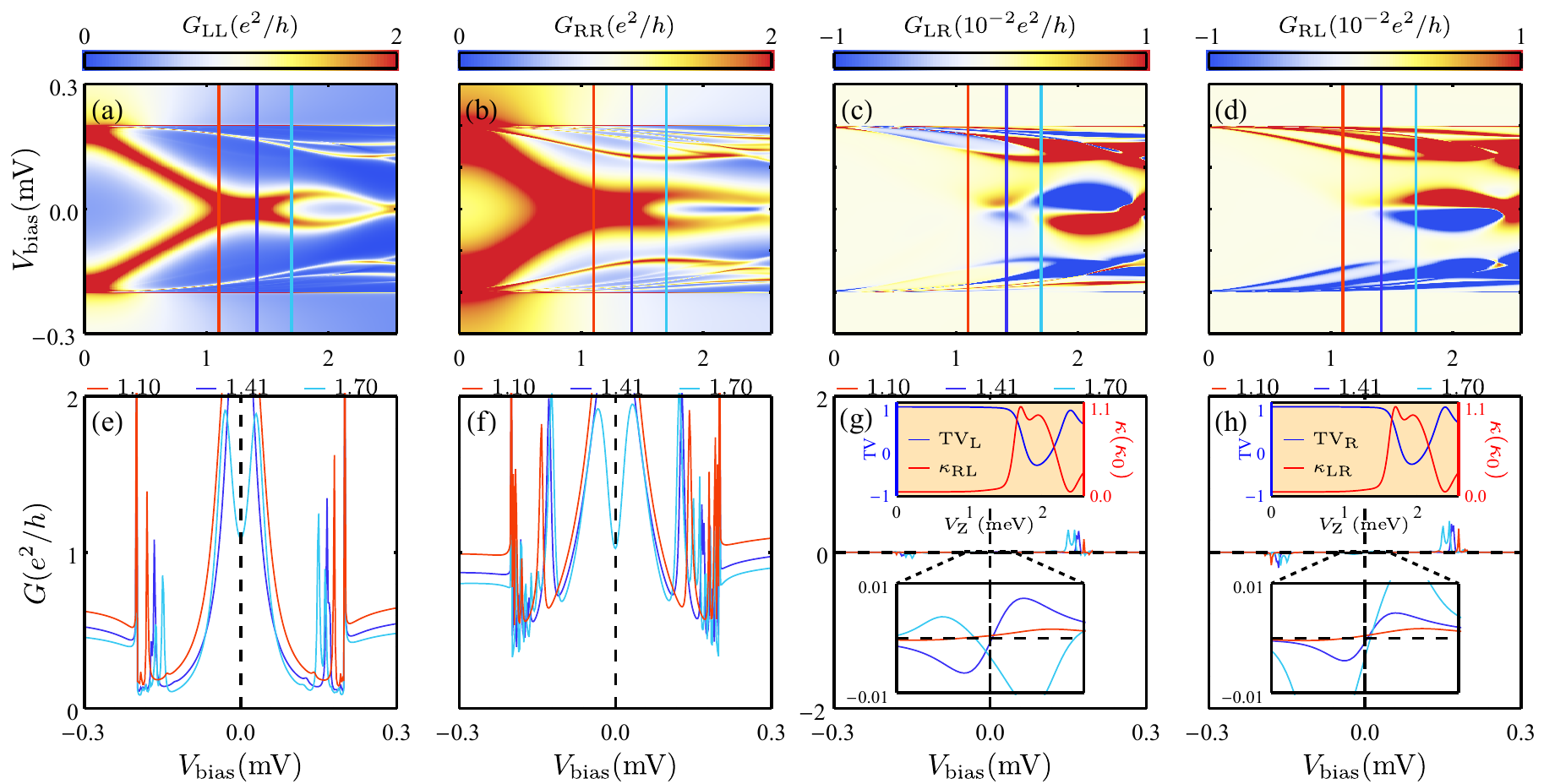}
	\caption{The short wire ($ L=0.5~\mu $m) in the presence of very strong disorder with $ \sigma_\mu/\mu=5 $. (a)-(d) show the local and nonlocal conductance in the ``intrinsic'' color scale. (e)-(h) are the corresponding line cuts of the conductance as a function of bias at $ V_{\text{Z}}=1.1 $ meV, $ 1.41 $ meV, $ 1.7 $ meV. The other parameters are the same as Fig.~\ref{fig:good}. The corresponding TV from the left (right) and thermal conductance $ \kappa_{RL} $ ($ \kappa_{LR} $) are shown in the inset of (g) [(h)].}
	\label{fig:muVar5S}
\end{figure}

	In this section, we show other examples of the good, bad, and ugly ZBCPs in Figs.~\ref{fig:muVar2}-\ref{fig:muVar5S}.

\end{document}